\documentclass[reprint,superscriptaddress,amssymb, nobibnotes, aps, prd]{revtex4-2}
\usepackage[normalem]{ulem}
\usepackage[T1]{fontenc}
\usepackage{amsmath}
\usepackage{soul}
\usepackage[hidelinks]{hyperref}
\usepackage{tikz}
\graphicspath{{figures/}}
\usetikzlibrary{quantikz2}

\newcommand{\bH}{\hat{H}} 
\newcommand{\ba}{\hat{a}} 
\newcommand{\bb}{\hat{b}} 
\newcommand{\bphi}{\hat{\varphi}}

\begin{document}
\title{Quantum non-demolition measurements as a practical primitive for fault-tolerant computation against biased noise}%

\author{Christophe Vuillot}%
\email{christophe.vuillot@alice-bob.com}
\affiliation{Alice \& Bob, 49 Bd du Général Martial Valin, 75015 Paris, France}
\author{Diego Ruiz}%
\email{diego.ruiz@alice-bob.com}
\affiliation{Alice \& Bob, 49 Bd du Général Martial Valin, 75015 Paris, France}
\affiliation{Laboratoire de Physique de l'École Normale Supérieure, École Normale Supérieure, Centre Automatique et Systèmes, Mines Paris, Université PSL, CNRS, Inria, Paris, France}
\author{Jérémie Guillaud}%
\email{jeremie.guillaud@alice-bob.com}
\affiliation{Alice \& Bob, 49 Bd du Général Martial Valin, 75015 Paris, France}
\author{Mazyar Mirrahimi}%
\email{mazyar.mirrahimi@inria.fr}
\affiliation{Laboratoire de Physique de l'École Normale Supérieure, École Normale Supérieure, Centre Automatique et Systèmes, Mines Paris, Université PSL, CNRS, Inria, Paris, France}

\begin{abstract}
Leveraging noise bias, where phase-flip errors dominate over bit-flips, can drastically reduce the hardware overhead of fault-tolerant quantum computation, but existing approaches require bias-preserving CNOT gates whose implementation remains experimentally challenging and is provably impossible for strictly two-dimensional systems. 
We show that high-fidelity quantum non-demolition (QND) multi-qubit Pauli $Z$ measurements provide an equally powerful yet more accessible primitive. 
We demonstrate that such measurements can fully replace bias-preserving CNOT gates for compiling all operations required by bias-tailored error correction, including stabilizer measurements for repetition codes, XZZX surface codes, and LDPC codes.
We propose concrete physical implementations of this primitive for two platforms: solid-state nuclear spins coupled to electron spin ancillas, and dissipatively stabilized superconducting cat qubits. Through circuit-level numerical simulations, we show that an asymmetric XZZX surface code implemented with weight-four QND $Z$ measurements achieves a phase-flip threshold of $\sim\!1.25\%$ and provides a qubit overhead reduction of up to $6\times$ compared to a bias-unaware surface code at noise bias $\eta = 10^4$.
In the regime of very large bias, a repetition code with QND $Z$ measurements attains a threshold of $\sim\!2.3\%$ and achieves overhead comparable to that of a bias-preserving CNOT scheme, without requiring such a gate. 
Our results establish QND multi-$Z$ measurements as a practical and hardware-efficient route to fault-tolerant quantum computation for a broad class of biased-noise platforms.
\end{abstract}

\maketitle

\section{Introduction}

The promise of reliably simulating complex many-body quantum systems, as well as performing certain computations exponentially faster, has created a lot of enthusiasm around the implementation of quantum processors. Circumventing the impact of noise on such processors is a daunting problem that has concentrated  a major part of the research effort throughout the past three decades. The theory of quantum error correction and fault-tolerant computation provides the route towards handling  noise but comes at the expense of an increased complexity in the physical implementation of such processors \cite{shor1996ftqc,campbell2017roads}.  In parallel with the impressive experimental progress in demonstrating quantum error correction and the first steps towards realizing fault-tolerant quantum operations \cite{bluvstein2023logical, acharya2024surfacecode}, an important body of theoretical research is focused on reducing the hardware overhead of fault-tolerance. 

One promising approach towards such hardware-efficient fault-tolerant processors is to take advantage of the noise bias that is either naturally present in some physical platforms or is engineered through a built-in protection against one type of noise \cite{aliferis2008biased, tuckett2018ultrahigh, guillaud2019repetition}. Indeed, generally, the noise can alter the state of a quantum bit in two ways, either by causing an undesired and/or random transition between the states of the computational basis $\{\ket{0},\ket{1}\}$, or by scrambling the phase in the  superposition of these states. These two effects are however usually due to very different noise mechanisms. In most physical realizations (such as atomic or spin qubits), the first effect is due to a relaxation mechanism through an exchange of energy with the environment and its rate is determined by high-frequency components of noise. The second dephasing effect is however mainly due to an entanglement with the environment degrees of freedom and can also occur due to low-frequency components of noise. This major difference in the sources of the two effects usually leads to a large bias where the phase-flip rate significantly exceeds the bit-flip one.  More importantly, even in systems where such a noise bias is not significant in practice (such as superconducting artificial atoms), one can engineer an autonomous protection against one type of error and effectively obtain a biased-noise qubit.  This is the case of cat qubits confined either through a Hamiltonian approach \cite{Puri-2017,grimm2020kerrcat} or a dissipative one \cite{Mirrahimi_NJP_2014,leghtas2015manifold, lescanne2020exponential}, noting that the latter leads to biases that are drastically larger than even the naturally noise-biased physical qubits, e.g noise bias of $10^8$ measured in recent experiments \cite{reglade2024catcontrol,rousseau2025}. 

The hardware resource savings that are enabled by the consideration of such a noise bias greatly depends on the set of the elementary physical operations that preserve this bias. Indeed, two approaches have been prominently explored so far. One is based on C\(Z\) gates that can be naturally bias-preserving \cite{aliferis2008biased,brooks-preskill-2013} and the other assumes bias-preserving C\(X\) gates that require physical systems and specific control enabling this particular gate \cite{guillaud2019repetition, puri2020biaspreserving}.
The first approach features a quite modest gain, mainly in the accuracy threshold value of the error correction. However, it requires rather complex gadgets to perform important and recurring operations such as the parity checks required for the error correction. The overall savings in the hardware overhead of fault\nobreakdash-tolerance appears to be very slim at best~\cite{Ruiz2026_Review}. The second approach unleashes the real power of noise bias to achieve fault\nobreakdash-tolerance. Indeed, a bias-preserving C\(X\) operation enables the use of error correcting codes that benefit from this error bias, either to significantly improve the error threshold~\cite{tuckett2018ultrahigh,bonilla2021xzzx}, or to reduce the number of physical qubits encoding a logical qubit by asymmetrically reducing the distance for one type of error~\cite{guillaud2019repetition,AWS-Blueprint-2022,shanahan2026}. In the case of extreme noise biases demonstrated for dissipative cat qubits, the cost of  error correction for a fault-tolerant application-scale quantum (FASQ) machine~\cite{eisert2025} could even be drastically reduced by relying on classical Low-Density Parity Check codes with a 2D local architecture~\cite{Ruiz-LDPC-2025}. Finally, more recently, we  showed that with such bias-preserving  C\(X\)  operations, even in case of more modest noise bias, we can expect drastic cost reduction for distillation of magic states required for fault-tolerant operations~\cite{Ruiz-npj-2026}. 

In all these exciting results, the bias-preserving operation of C\(X\) enables us to fully exploit the suppression of one type of error without requiring complex gadgets to perform recurring operations such as parity checks. However, on the downside, the implementation of such a bias-preserving C\(X\) gate is a challenging task on its own. First,  it can be demonstrated that such a C\(X\)  operation cannot be bias-preserving when performed on physical qubits defined strictly in a two-dimensional Hilbert space~\cite{guillaud2019repetition}. This therefore excludes some naturally biased-noise qubits such as electron spins. On the contrary, bosonic cat qubits~\cite{LesHouches-2023} or large nuclear spin cats~\cite{Gross-PRApp-2024,Marvian-PRXQ-2024,Kruckenhauser-PRL-2025,Morello-cats-2025} can in principle circumvent this no-go theorem through a carefully engineered interaction. Although various theoretical proposals indicate the possibility of implementing such an operation, a fully convincing experimental demonstration is still lacking. 

Interestingly, in a recent work~\cite{Claes-npj-2023}, the authors show that in the context of Measurement-Based Quantum Computation (MBQC)~\cite{Raussendorf-Briegel-2001}, it is possible to generate fault-tolerant cluster states which are tailored to benefit from the noise bias even in the absence of bias-preserving C\(X\) gates. Their approach relies on quantum non-demolition (QND) multi-qubit Pauli $Z$ measurements which can be performed through the application bias-preserving C\(Z\) gates between the data qubits and an ancilla which is then measured destructively along its $X$ axis. 
In this article, inspired by this observation, we show that high-fidelity QND multi-qubit $Z$ measurements can be used as a primitive for hardware-efficient and fault-tolerant quantum computation with biased-noise qubits. More precisely, we first demonstrate in Section~\ref{sec:primitive} that such measurements can fully replace bias-preserving C\(X\) gates with a small or no addition in hardware complexity. Second, in Section~\ref{sec:physical} we provide concrete proposals for performing these high-fidelity measurements in a hardware-efficient manner on physical platforms such as solid-state nuclear spins or stabilized superconducting cat qubits. Finally, in Section~\ref{sec:performances}, we numerically study the performance of various error correction approaches under physically relevant assumptions.    

\section{A different primitive for biased-noise quantum computation}\label{sec:primitive}

We note that the complexity of the bias-preserving C\(X\)  operation comes from the fact that, as a function of the state of the control qubit, it inverts the state of the target qubit in the computational basis. In a strictly two-dimensional system this boils down to a $\pi$-rotation around the qubit's $X$ axis (or any other axis orthogonal to $Z$), and a phase-flip during this rotation can actually be converted to a bit-flip type error at the end of the gate. In cat qubits this noise transformation is circumvented by inverting the computational states through an excursion in the large Hilbert space while always keeping these states far from each other in the bosonic mode's phase space. However, this comes at the expense of  modifying the confinement mechanism such that the confined states make the required excursion in the phase space. 

\begin{figure}[ht]
    \centering
    \begin{quantikz}[row sep=0.5em]
        &\ctrl{1}&\\
        \lstick{\ket{\Psi_0}} & \gate{U}&\meterD{M_{\left\{\ket{\Psi_0}, \ket{\Psi_1}\right\}}}
    \end{quantikz}
    \caption{Making a reliable QND \(Z\) measurement amounts to having access to a C\(U\) gate, a preparation of a state \(\ket{\Psi_0}\) and a reliable measurement distinguishing it from \(\ket{\Psi_1}=U\ket{\Psi_0}\).
    }
    \label{fig:CU}
\end{figure}
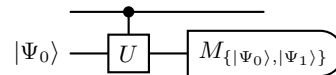

In contrast, for a high-fidelity QND measurement of Pauli $Z$ operators, we can rely on  a controlled-U operation between the biased noise qubit and a measurement device modeled as a quantum system prepared in initial state $\ket{\psi_0}$ such that $\ket{\psi_1}=U\ket{\psi_{0}}$ is macroscopically distinguishable from the state $\ket{\psi_0}$. Throughout this letter, we refer to this measurement device as the meter. The biased-noise qubit is only used as the control qubit and does not undergo any population transfer. This leads to a high-fidelity measurement of the biased-noise qubit in its $Z$ basis as depicted in Figure~\ref{fig:CU}. In principle, this operation can be done so that any error of the meter before or during the controlled-U gate only propagates to $Z$ type errors of the biased noise qubit. Furthermore, the fidelity of the measurement is determined by the distinguishability of the two states $\ket{\psi_0}$ and $\ket{\psi_1}$. Indeed, the assignment errors of the QND measurement  {play precisely the same role as} the bit-flip errors of the bias-preserving  C\(X\) operation {in a parity-check circuit for quantum error correction. We thus} need their rate to be orders of magnitude smaller than the phase-flip rate. 

\begin{figure}[ht]
    \centering
    \includegraphics[width=.49\textwidth]{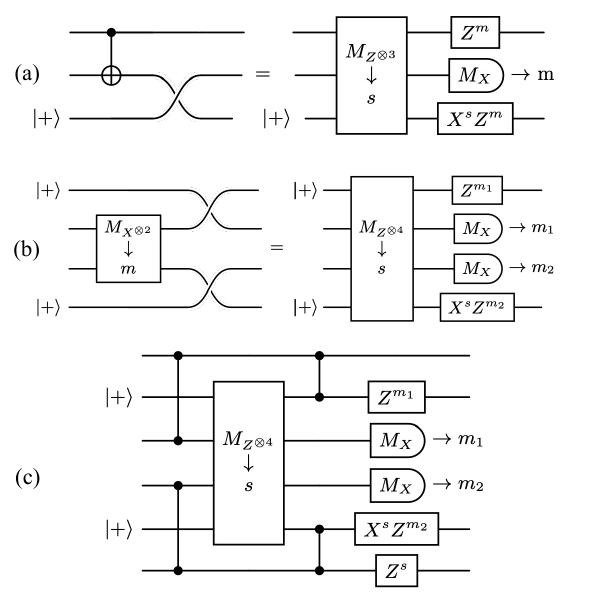}

    \caption{(a) Circuit to implement a C\(X\) gate using a QND \(Z^{\otimes 3}\) measurement. Depending on the measurement outcome there is an \(X\) correction to be applied. In absence of a bias-preserving \(X\) gate, this correction can be stored in a Pauli frame. Moreover, the assignment error of the \(Z^{\otimes 3}\) measurement has to be as small as the bit-flip error probability for the protocol to preserve the bias. (b) Circuit to implement an \(X^{\otimes 2}\) parity check using a single QND \(Z^{\otimes 4}\) measurement. The measurement outcome is given by \(m=m_1\oplus m_2\). (c) Circuit to implement a \(ZXXZ\) measurement using a single QND \(Z^{\otimes 4}\) measurement. The measurement outcome is given by \(m=m_1\oplus m_2\). It can be obtained by conjugating the \(XX\) measurement in plot (b) by C\(Z\) gates.}
    \label{fig:CX}
\end{figure}

Now, let us argue why a high-fidelity QND \(Z\)-type measurement is as powerful as a bias preserving C\(X\) gate to ensure hardware-efficient error correction and fault-tolerance.
Mainly, as shown in Fig.~\ref{fig:CX}(a), one can implement a bias preserving C\(X\) gate up to a Pauli \(X\) correction that can be stored in a Pauli frame. This implementation uses an extra qubit prepared in the \(\ket{+}\) state, performs a QND \(Z^{\otimes 3}\) measurement and then an \(X\) readout.
In the final state there is an \(X\) correction depending on the first \(Z\)\nobreakdash-type measurement outcome that is either kept in a Pauli frame or can be applied if the physical system allows a bias preserving implementation of \(X\). Importantly this implies that for this protocol to be bias-preserving, the assignment error of the QND \(Z\)\nobreakdash-type measurement outcome needs to be as small as the bit\nobreakdash-flip error probability, besides all non\nobreakdash-QND type errors also respecting the bias. In contrast, the result of the final $X$\nobreakdash-measurement on the first qubit determines a feedforward \(Z\) operation on output qubits. Thus, the assignment errors of this measurement lead to correlated phase\nobreakdash-flip errors. Finally, we also note that the errors of ancilla preparation as well as any phase-flip errors during the QND \(Z^{\otimes 3}\)\nobreakdash-measurement propagate through as phase\nobreakdash-flip errors.  Another particularity of this protocol is that the extra qubit is permuted with the two initial qubits during the protocol, see Fig.~\ref{fig:CX}(a).
More precisely the target qubit in the C\(X\) always ends up on the extra qubit and the control can be chosen on either of the two other initial qubits by choosing the other one to be measured in \(X\). Note that while this protocol can already be found in \cite{aliferis2008biased,brooks-preskill-2013},  it is only used at the logical level of a repetition code, which requires quite a complex and costly gadget (see Appendix~\ref{sec:CZ}). Here we propose to use it at the physical level directly.

With this simple observation, one can realize all proposals that rely on bias-preserving C\(X\) gates using such QND \(Z\)\nobreakdash-type measurements. Moreover,  logical circuits based on such C\(X\) gates can be re-compiled using QND multi-qubit \(Z\) measurements, while minimizing  the additional complexity. Performing error correction in the context of biased noise typically involves measuring \(X\)\nobreakdash-type or mixed-type (e.g. \(XZZX\)) Pauli stabilizers. Using the C\(X\) protocol of the previous section one can compile such Pauli measurements with \(X\) components in the measured Pauli operator in a straightforward manner. It is also possible to reduce the depth and number of QND \(Z\)\nobreakdash-type measurements at the cost of a higher weight QND \(Z\)-type measurement. For instance, Figure~\ref{fig:CX}(b) describes a weight\nobreakdash-two \(X\) measurement using a single weight-four QND \(Z\) measurement and two auxiliary qubits. This generalizes straightforwardly to implementing a weight\nobreakdash-\(k\) \(X\) measurement using a weight\nobreakdash-\(2k\) QND \(Z\) measurement and \(k\) ancilla qubits. We also show in Appendix~\ref{sec:ldpc_parity} how to compile weight\nobreakdash-\(k\) \(X\) measurements using only weight\nobreakdash-four QND \(Z\) measurements.
From \(X\)-type measurements it is direct to obtain mixed \(XZZX\)-type measurement by using C\(Z\) gate conjugating the measurement as shown in Figure~\ref{fig:CX}(c). 

\section{Physical implementation}\label{sec:physical}

One possible way to realize such QND and high-fidelity multi-Z Pauli measurements is to rely on bias-preserving C\(Z\) gates and use a meter composed of $N$ ancilla qubits all prepared in the state $\ket{+}$. In order to perform a $Z^{\otimes q}$\nobreakdash-measurement with $q$ data qubits, we perform $N\times q$  C\(Z\) gates between the data qubits and each of ancilla qubits and finally measure each of the ancilla qubits in its $X$ basis. The high-fidelity measurement result is given by a majority vote on the results of all $N$ measurements. Note that by increasing the number of ancilla qubits $N$, the multi-Z assignment error probability decreases exponentially. This effectively means that the bit-flip error probability in the above protocols decreases exponentially with $N$. However, the measurement circuit volume increasing linearly with $N$, this boils down to a linear increase in the phase-flip error probability. 

While one might find some resemblance between such a protocol and the C\(Z\)-based gadgets for biased-noise qubits proposed in~\cite{aliferis2008biased,brooks-preskill-2013}, the crucial difference is that the protocols in these references rely on multi-Z measurements at the logical level of a repetition code composed of biased noise qubits. Here, instead, such multi-Z measurements are performed directly at the physical level. This leads to a remarkable simplification of the gadgets and implies a significant performance improvement with respect to these early C\(Z\)-based protocols. The Appendix~\ref{sec:CZ} provides a quantitative comparison between the performances of these C\(Z\)-based protocols.

However, the real interest of the multi-Z measurement primitive for hardware savings in fault-tolerance  comes from the fact that the meter does not need to be composed of the same physical qubits as the ones used for computation. In the following subsections, we consider two particular physical settings and illustrate in each one how such a high-fidelity QND measurement can be done in a hardware-efficient manner. 

\subsection{Nuclear spins as biased noise qubits}\label{ssec:nuclear}
\begin{figure}[ht]
    \centering
    \includegraphics[width=.5\linewidth]{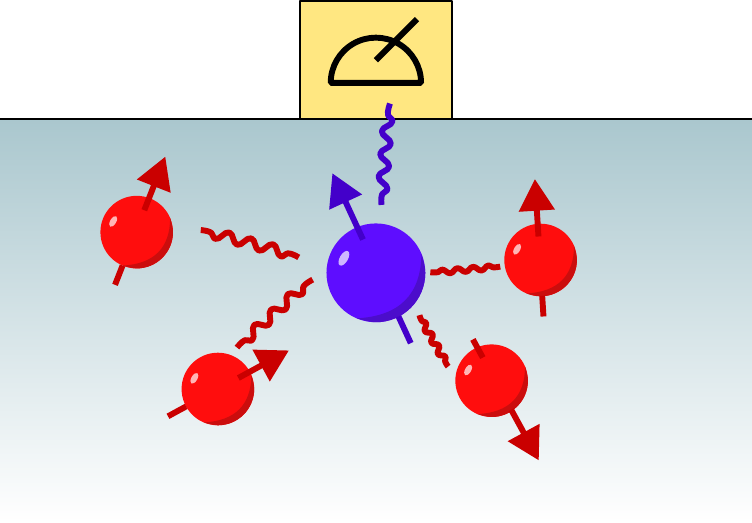}\\[1em]
    \caption{Four nuclear spins (in red) are coupled to an electron spin (in blue) through hyperfine interaction and the electron spin undergoes an optical or electrical readout. Such a setup could be used to perform high-fidelity QND measurement of the Pauli operator $Z^4$ between the four nuclear spin qubits which in turn can be used to perform the parity checks required for biased-noise error correction as in Fig.~\ref{fig:CX}.}
    \label{fig:spins}
\end{figure}

Consider the case of solid state nuclear spins coupled to electron spin ancillas, controlled and measured either optically~\cite{Taminiau-PRL-2012,Taminiau-NatComm-2016,Taminiau-Nature-2022} or electrically~\cite{Kane-Nature-98,Pla-Nature-2013,Morello-Nature-2022}, see Fig.~\ref{fig:spins}. In the particular case of experiments with  donors in Silicon, the high-fidelity and QND measurement of single nuclear spins have been demonstrated through a C\(X\) operation between the nuclear spin and the electron spins and through the single-shot readout of the electron spin by spin-to-charge conversion~\cite{Pla-Nature-2013}. This scheme can be transformed  into a multi-Z Pauli measurement on a few nuclear spin qubits by coupling them to a single electron spin as in~\cite{Morello-Nature-2022}. More precisely, let us consider the Hamiltonian of the system
$$
\bH=\bH_0+\bH_{\text{rf}}(t),
$$
with
\begin{equation}\label{eq:spinH0}
\bH_0=-\gamma_e B_0 \hat S_z-\gamma_n B_0 \sum_{k=1}^q \hat I_{k,z}+\sum_{k=1}^q A_{k}\hat S_z\hat I_{k,z},
\end{equation}
and
\begin{equation}\label{eq:spinHrf}
\bH_{rf}(t)=-\gamma_e B(t) \hat S_x,
\end{equation}
where $\gamma_e$ and $\gamma_n$ are electron and nuclear gyromagnetic ratios, $B_0$ corresponds to the static magnetic field aligned with the $Z$ axis, $\hat S $ and $\{\hat I_k\}_{k=1}^q$ are, respectively, the electron spin and nuclear spin  operators, $A_k$'s are the hyperfine interaction strengths, and $B(t)$ corresponds to an AC microwave drive which is used to control the electron spin state. Note that while this field would also drive the nuclear transitions, its effect can be neglected as the effective transition frequencies are very different.  Now, in order to map the multi-Z Pauli information of the nuclear spins to the electron's spin{, and in a similar manner to~\cite{OSullivan-NatPhys-2025},} one can apply a microwave drive of the form 
\begin{align}\label{eq:drive}
B(t)&=\bar B\sum_{\substack{r\in\{0,1\}^q;\\ \sum r_k=1 (\text{mod }2)}}\cos(\omega_r t),\notag\\ 
\omega_r&=\gamma_eB_0-\frac{1}{2}\sum_{k=1}^q (-1)^{r_k} A_k.
\end{align} 
After a time given by $T_{\text{pulse}}=2\pi/\gamma_e\bar B$, the multi-Z information is mapped to the electron's excitation.

 Note that while this mapping operation is not bias-preserving (as performed on two-level systems), nuclear spins do not undergo bit-flip errors as they are used as control qubits. Now, current  implementations of electron spin readout through spin-to-charge conversion could scramble the phase of the nuclear spin states through a sudden change of their hyperfine coupling to the electron spin. In order to avoid such a dephasing process, one can apply appropriate dynamical decoupling sequences during the spin-to-charge conversion and readout to protect the nuclear spin coherence~\cite{Wolfowicz_2016}. Furthermore, note that the same idea could be used in a number of other platforms with different implementations of electron spin readout. {Instead of measurement via localized electrons, it is possible to couple the nuclear spins to quantum dots~\cite{Hensen_Nat_Nano,Steinacker_PRL}. In this case, the strong inter-dot tunnel coupling rate with respect to the hyperfine coupling provides deterministic control over the movement of the electron charge,  thus avoiding a nondeterministic phase shift in the nuclear spins.}   In experiments with nitrogen-vacancy (NV) centers in diamonds, the optical readout of the electronic spin has been performed without scrambling the phase of the nearby $^{13}$C nuclear spins thus implementing a projective measurement of multi-spin systems~\cite{Jiang-PRL-2008,Pfaff-NatPhys-2013}. Further, it is also possible to instead map the multi-Z information of the nuclear spins to the excitation of a quantum dot which is then dispersively measured through its coupling to a superconducting resonator~\cite{Vandersypen-Science-2018}. Finally, in the context of recently developed all microwave readout of electron spins through their magnetic coupling to a detection microwave resonator~\cite{Travesedo-2025}, one should be able to implement such QND multi-Z measurement of neighboring nuclear spins through a dispersive readout of the electronic spin. 
 
 While the bit-flip errors of the electron spin ultimately limit the measurement fidelity, it is possible to reach very high fidelities by repeating this process many times. The multi-Z readout fidelity is thus ultimately determined by the deviations from the QND nature of these measurements. In the above experiments~\cite{Pla-Nature-2013,Morello-Nature-2022} such deviations are induced by phenomena such as the cross-relaxation of the nuclear spin through their transverse hyperfine couplings to electron spins or the ionization shock during the electron spin readout. However, it is possible to remove such parasitic processes, for instance, by applying larger static magnetic fields. {Indeed, as discussed in~\cite{Travesedo-2025}, the cross relaxation rate decreases as $B_0^{-2}$.}  It is thus plausible to expect very high fidelity multi-Z Pauli measurements with assignment error probabilities only limited by the natural nuclear spin relaxation. This assignment error probability ultimately sets the bit-flip error probability of the above protocols, which will thus be given by the ratio between the high-fidelity multi-Z measurement duration and the nuclear spin relaxation time.

On the other hand, the phase-flip error probability is set by two things: 1- the decoherence of the nuclear spins during the whole process of mapping and electron readout, 2- the decoherence of the electron spin during the mapping stage as it can propagate to the nuclear spins. Noting that such decoherence in spin systems is mainly due to low-frequency noise components, the second contribution can be made  small through fast adiabatic passage techniques. Thus, the phase-flip error probability is ultimately set by the ratio between the duration of the multi-Z Pauli readout and the nuclear spin coherence time. 

{Before switching to the case of superconducting cat qubits, let us mention that these ideas can, in principle, be transferred to other quantum computing platforms such as tweezer arrays of cold atoms or trapped ions. }

\subsection{Superconducting cat qubits}\label{ssec:cats}

After this brief description of an implementation with naturally biased-noise spin systems, let us  describe how the proposal of this article also leads to an alternative roadmap for operating biased-noise bosonic cat qubits. As mentioned before, the dissipatively stabilized cat qubits benefit from extremely high noise biases. Furthermore, theoretically, one should be able to perform  C\(X\) operations while preserving this noise bias. However, the implementation of such a bias-preserving C\(X\) gate is not completely straight-forward and requires further ingredients such as the  realization of new interaction Hamiltonians and/or adiabatic variation of parametric drivings. Except for some preliminary results~\cite{Cortinas-APS-2021,Essig-APS-2023}, a convincing experimental demonstration of such a gate is still missing both in the framework of dissipative confinement of cat qubits and their Hamiltonian Kerr confinement. The continuous QND measurement of the cat qubit along its \(Z\) axis is, however, demonstrated in both cases~\cite{grimm2020kerrcat,reglade2024catcontrol}. 

\begin{figure}[ht]
    \centering
    (a)\includegraphics[width=0.7\linewidth]{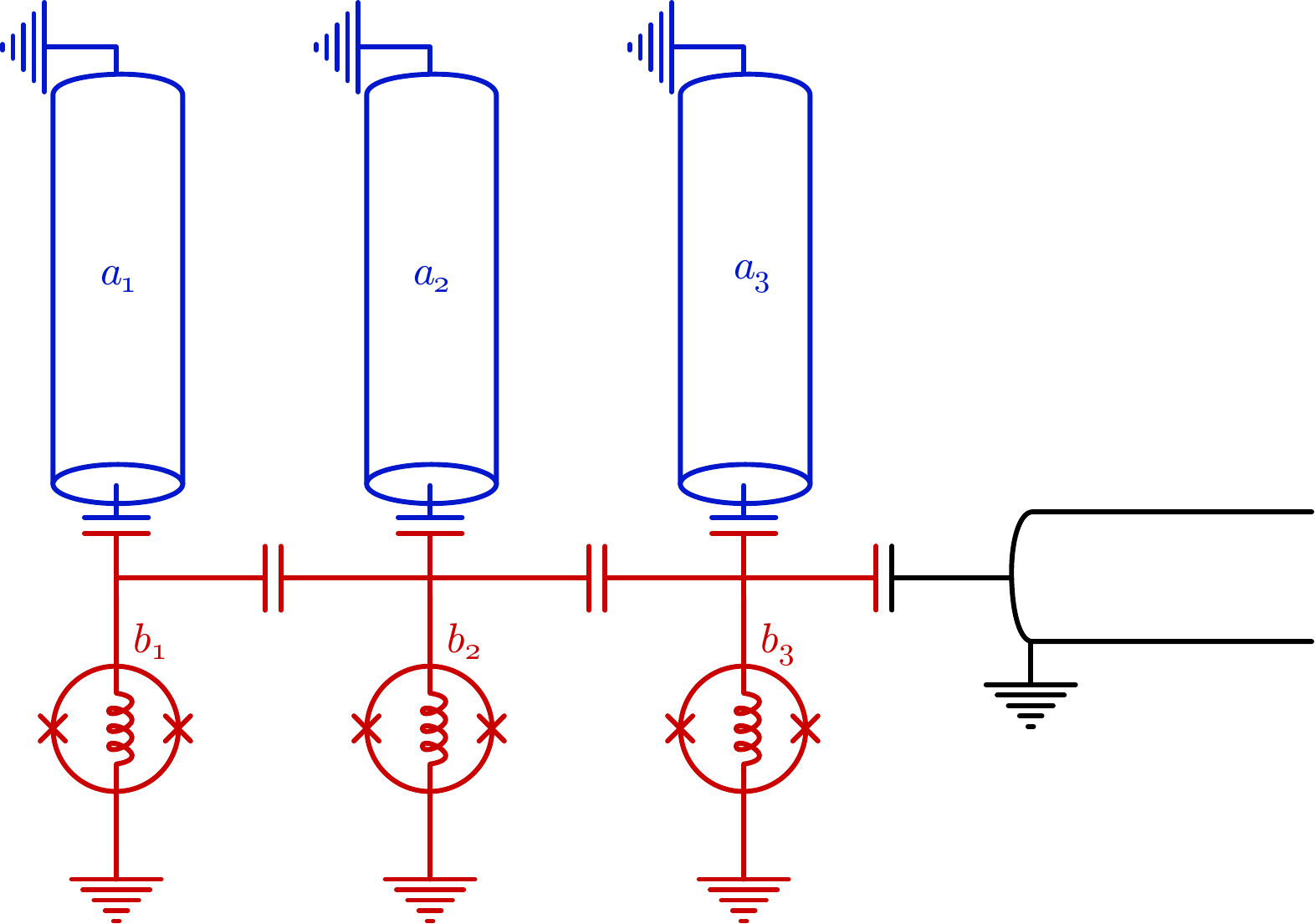}\\[1em]
    (b) \includegraphics[width=0.7\linewidth]{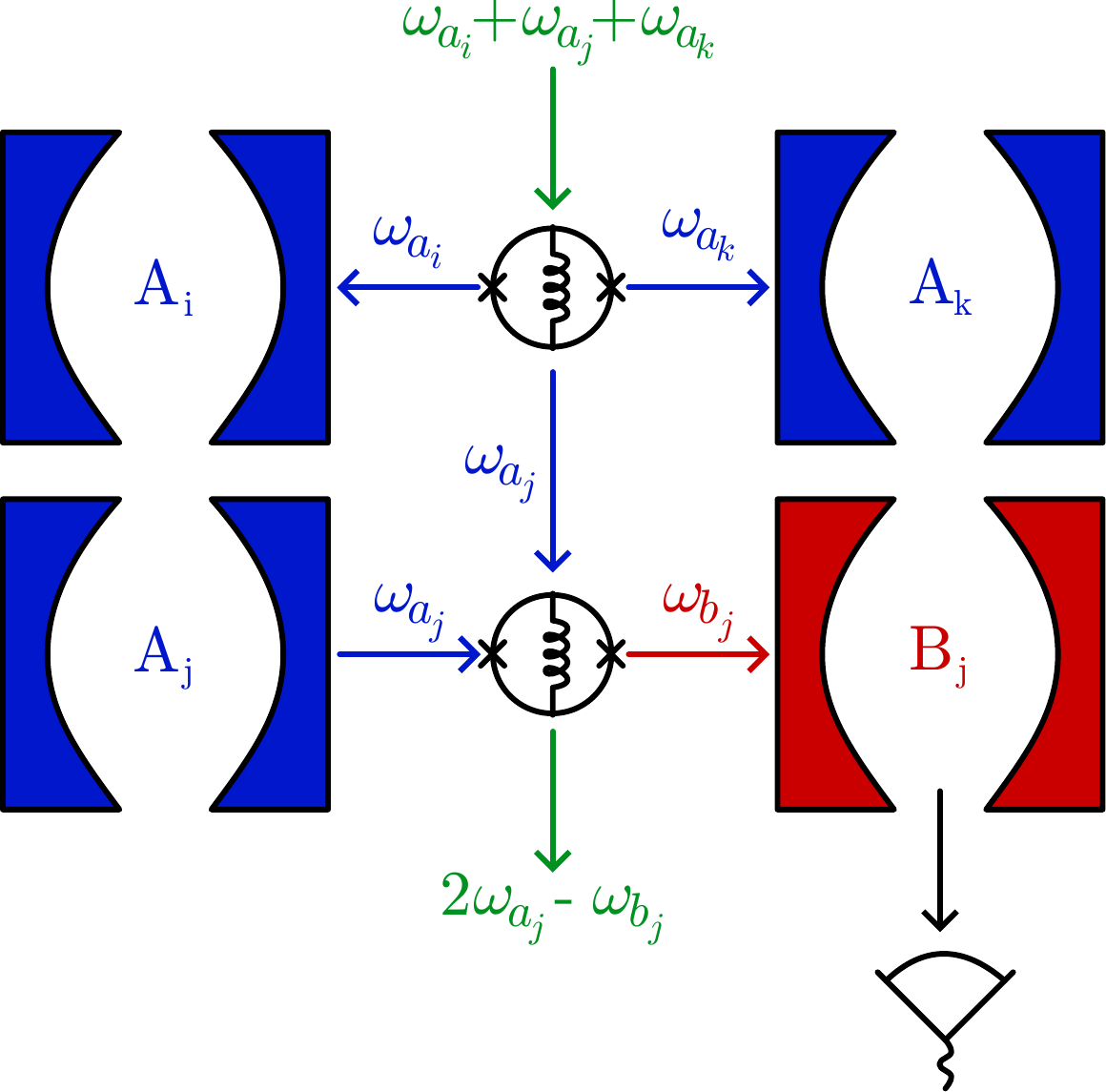}
    \caption{High-fidelity QND readout of $Z^{\otimes 3}$ between three stabilized cat qubits. As shown in plot (a), the setup consists of three dissipatively stabilized cat qubits coupled capacitively. Each cat qubit mode $a_j$ (in blue) is coupled to its buffer mode $b_j$ (in red). The Asymmetrically Threaded SQUIDs (ATS), represented by SQUIDs shunted in the middle by an inductance, ensure the nonlinear interactions between the cat qubit modes and the buffer modes.  By applying  dc magnetic fluxes of 0 and $\pi$ in the two superconducting loops of the ATS, this element engineers a potential of the form~\eqref{eq:ATSpot}. The $j$'th ATS is pumped with an ac drive at frequency $2\omega_{a_j}-\omega_{b_j}$, ensuring effectively an exchange of two photons at the memory frequency $\omega_{a_j}$ to a single photon at the lossy buffer frequency $\omega_{b_j}$. While such a process in presence of a resonant buffer drive stabilizes a cat qubit in the mode $\hat a_j$, we also apply on one or all of these ATS's a pump at frequency $\sum_k \omega_{a_k}$. As shown in plot (b), this pump combines with the pump at frequency $2\omega_{a_j}-\omega_{b_j}$ in a second order process, to convert a photon at frequency $\omega_{a_j}$ to three photons at frequencies $\omega_{a_i}$, $\omega_{a_k}$ and $\omega_{b_j}$. As explained in the main text, the effect of this conversion, in the presence of cat qubit stabilization, is to map the multi-qubit Pauli $Z$ information of the three cat qubits to the phase of the coherent field leaving the buffer mode $\hat b_j$. A heterodyne detection of the outgoing field thus leads to a QND readout of the multi-qubit Pauli Z operator.   }
    \label{fig:cats}
\end{figure}

Indeed, in the case of Kerr cat qubits, the experiment~\cite{grimm2020kerrcat} and its recent improvements~\cite{Frattini-PRX-2024,Qing-PNAS-2026,Grimm-Arxiv-2025} performed such a QND measurement by realizing a beam-splitter interaction of the form 
\begin{equation}\label{eq:BS}
\hat H= i\hbar g_{cr}(\hat a\hat b^\dag-\hat a^\dag\hat b)
\end{equation}
between the cat qubit mode (annihilation operator $\hat a$) and a readout cavity mode (annihilation operator $\hat b$).  In combination with the confinement of the cat qubit mode in $\text{span}\{\ket{\alpha},\ket{-\alpha}\}$ (we assume $\alpha$ real, in order to simplify the notations), the above Hamiltonian effectively realizes the interaction $\hat H_{bs}=i\hbar \alpha g_{cr}\hat Z\otimes (\hat b^\dag-\hat b)$ where $\hat Z$ represents the Pauli $Z$ operator of the cat qubit (note that in~\cite{grimm2020kerrcat}, the convention of the authors was to call this axis $X$).  In all these experiments the fidelity of the QND measurement is mainly limited by the same processes limiting the bit-flip time of the confined cat qubit. In the case of the dissipative cat qubits a similar QND readout can be engineered through the addition of a beam-splitter interaction with a readout mode, and should reach much higher fidelities due to better protection against bit-flips. In this dissipative case, it is also possible to simply readout the cat qubit through the buffer mode that is used for the two-photon dissipation with the addition of a resonant drive of appropriate phase on the cat qubit mode as shown in~\cite{reglade2024catcontrol,rousseau2025}. Throughout the next few paragraphs, we will extend this readout scheme to a high-fidelity QND multi-Z readout.

By considering the equivalent circuit of the C\(X\) gate  in Fig.~\ref{fig:CX}(a), we focus on weight-3 Pauli \(Z\) measurements as all other logical circuits can be compiled by combining such C\(X\) gates. Let us recall that the stabilization of the cat qubits is ensured through a two-photon exchange process with a driven damped buffer mode as modeled by the following Hamiltonian~\cite{Mirrahimi_NJP_2014}:
\begin{equation}\label{eq:H2ph}
\bH_{2\text{ph}}=\hbar\sum_{k=1,2,3} g_{2\text{ph},k}((\hat a_k^2-\alpha_k^2)\hat b_k^\dag+\text{h.c.})
\end{equation}
 where $\hat a_k$ corresponds to the annihilation operator for the bosonic mode of the $k$'s cat qubit, $\hat b_k$  corresponds to the associated buffer mode, and $g_{2\text{ph},k}$ the engineered two-photon exchange strength, assumed real to simplify the notations. The buffer modes being lossy, this engineered interaction Hamiltonian leads to effective loss channels given by $L_k=(\hat a_k^2-\alpha_k^2)$ and stabilizing the cat qubits $\text{span}\{\ket{\pm\alpha_k}\}$. Such an interaction is in practice engineered through the circuit of Fig.~\ref{fig:cats}(a). Each cat qubit mode $\ba_k$ is coupled to its buffer mode $\bb_k$, through a nonlinear element called the ATS (Asymmetrically Threaded SQUID)~\cite{lescanne2020exponential}. By applying  dc magnetic fields in the two superconducting loops of the ATS, we set the reduced fluxes to be 0 and $\pi$. The $k$'th ATS element then effectively implements a potential of the form
\begin{equation}\label{eq:ATSpot}
U_k(\bphi_k)=-2E_{J,k}\epsilon_k(t)\sin(\bphi_k),
\end{equation}
where $E_{J,k}$ represents the Josephson energy of the junctions of the $k$'th ATS, $\epsilon_k(t)$ is the ac  flux pump applied to the ATS, and $\bphi_k$ is the phase drop across that ATS. Here this phase drop is given by a linear combination of all field mode operators
$$
\bphi_k=\sum_{j=1,2,3} \varphi^a_{k,j}(\ba_j+\ba_j^\dag)+\varphi^b_{k,j}(\bb_j+\bb_j^\dag).
$$
To engineer an effective interaction of the form~\eqref{eq:H2ph} with $\alpha_k=0$, we choose $\epsilon_k(t)=\bar\epsilon_k \cos(\omega_{p_k} t)$ with $\omega_{p_k}=2\omega_{a_k}-\omega_{b_k}$, where  $\omega_{a_k}$ and $\omega_{b_k}$ represent the resonance frequencies of the modes $\ba_k$ and $\bb_k$. The interaction strengths $g_{2\text{ph},k}$ are given by 
$g_{2\text{ph},k}=E_J\bar\epsilon_k (\varphi_{k,k}^a)^2\varphi_{k,k}^b/\hbar$. Furthermore, by applying a resonant drive on the buffer mode $\bb_k$, we can now vary $\alpha_k$ and therefore the average photon number in the stabilized cat qubits. The efficient stabilization of cat qubits through such an approach has been demonstrated in many recent experiments~\cite{lescanne2020exponential,reglade2024catcontrol,rousseau2025}. Note furthermore that, it is also possible to engineer such 3-wave mixing process for stabilization of cat qubits through a resonant interaction, without the requirement of a parametric driving~\cite{Marquet-PRX-2024}.

Now, to ensure a multi-Z QND readout for these 3 coupled cat qubits, we propose to apply an additional flux pump on any of the ATS elements. More precisely, we propose to consider a flux pump of the form
$$
\epsilon_k(t)=\bar\epsilon_{k}\cos(\omega_{p_k} t)+\bar\epsilon_\Sigma\cos(\omega_{\Sigma} t)
$$
with  $\omega_{\Sigma}=\sum_{j} \omega_{a_j}$. The ATS, pumped at this second  tone, engineers a new 3-wave mixing term of the form
\begin{equation}\label{eq:H3}
\bH_{\text{int}}=\hbar g_{Z} (\hat a_1\hat a_2\hat a_3+\text{h.c.}),
\end{equation} 
with the interaction strength $g_{Z}=2E_J\bar\epsilon_\Sigma\varphi_{k,1}^a\varphi_{k,2}^a\varphi_{k,3}^a/\hbar$.
The particular form of the above interaction is chosen for symmetry reasons, but one could instead engineer an interaction of the form $\hbar g_Z(\hat a_1\hat a_2\hat a_3^\dag+\text{h.c.})$ where the frequency matching conditions are perhaps easier to achieve. The analysis below can be adapted in a straight-forward manner to such an interaction. 
 
 Let us now study the equations of motion under the Hamiltonian $\bH=\bH_{2\text{ph}}+\bH_{\text{int}}$ and the loss of the buffer modes $\hat b_k$. The Langevin equations of motion are given by
 \begin{align}\label{eq:langevin}
 \frac{d}{dt}\hat a_1&=-2ig_{2\text{ph},1}\hat a_1^\dag \hat b_1-ig_Z \ba_2^\dag \ba_3^\dag ,\\
 \frac{d}{dt}\hat b_1&=-ig_{2\text{ph},1}(\hat a_1^2-\alpha_1^2)-\frac{\gamma_1}{2}\hat b_1+\sqrt{\gamma_1}\bb_{1,\text{in}},\notag
 \end{align}
 and similar equations for modes 2 and 3, where $\gamma_k$ represents the decay rate of the $k$'th buffer mode. Furthermore, the  buffer mode's output field is given by $\hat b_{k,\text{out}}=\hat b_{k,\text{in}}+\sqrt{\gamma_k}\hat b_k$. Assuming a vacuum input, we show  that the heterodyne  readout of the output field of the buffer modes,  effectively measures the multi-Z Pauli operator associated with the cat qubits. 

We  linearize the above Langevin's equations around $(\hat a_k,\hat b_k)=(\alpha_k Z_k,0)$ where $Z_k$ stands for the value of the Pauli Z on the $k$'s cat qubit's. More precisely, by writing $(\hat a_k,\hat b_k)=(\alpha_k Z_k,0)+(\widehat{\delta a}_k,\widehat{\delta b}_k)$, and keeping the first-order terms in $(\widehat{\delta a}_k,\widehat{\delta b}_k)$, we obtain for modes 1 (similar equations for modes 2 and 3)
 \begin{align}\label{eq:linearized}
 \frac{d}{dt}\widehat{\delta a}_1=&-2ig_{2\text{ph},1}\alpha_1 Z_1\widehat{\delta b}_1-i g_Z \alpha_1^2 Z_2Z_3 \notag\\
 &-i g_Z(\alpha_2 Z_2\widehat{\delta a}_3^\dag+\alpha_3Z_3\widehat{\delta a}_2^\dag)\\
 \frac{d}{dt}\widehat{\delta b}_1=&-2ig_{2\text{ph},1}\alpha_1 Z_1\widehat{\delta a}_1-\frac{\gamma_1}{2}\widehat {\delta b}_1,\notag
 \end{align}
 where $\alpha_k$'s are taken to be real. The steady state solution of the above linearized system satisfies (similar equations for modes 2 and 3)
 \begin{align}\label{eq:ss}
  \widehat{\delta a}_1&=i\frac{\gamma_1}{4\alpha_1 g_{2\text{ph},1}}Z_1\widehat{\delta b}_1,\\
   \widehat{\delta b}_1&=-\frac{\alpha_1g_Z}{2g_{2\text{ph},1}}Z_1Z_2Z_3\notag\\
   &+i\frac{g_Z}{8\alpha_1g_{2\text{ph},1}}Z_1Z_2Z_3\left(\frac{\alpha_3\gamma_2}{\alpha_2 g_{2\text{ph},2}}\widehat{\delta b}_2^\dag+\frac{\alpha_2\gamma_3}{\alpha_3g_{2\text{ph},3}}\widehat{\delta b}_3^\dag \right).\notag
 \end{align}
 Thus a heterodyne readout of the buffer modes output field directly measures the multi-Z Pauli operator on the 3 coupled cat qubits. Assuming, without loss of generality, that the three cat qubits have the same parameters ($g_{2\text{ph},k}=g_{2\text{ph}}$, $\alpha_k=\alpha$, $\gamma_k=\gamma$), and by adding the output field signals, we achieve
$$
\hat{b}_{\Sigma}=\frac{1}{\sqrt{3}}\sum_{k=1,2,3}\widehat{\delta b}_k=-\frac{\zeta}{1-\nu^2}(Z_1Z_2Z_3+i\nu),
$$
with
$$
\zeta=\frac{\sqrt{3}\alpha g_Z}{2g_{2\text{ph}}},\quad \nu=\frac{g_Z\gamma}{4\alpha g_{2\text{ph}}^2}.
$$
This readout is performed while the cat qubits are autonomously protected against bit-flips through the two-photon driven dissipation mechanism. 

For $\nu\ll 1$, the buffer modes remain close to coherent states and therefore the measurement rate is approximated by
$$
\Gamma_{\text{meas}}\approx 2\gamma\zeta^2=\frac{3\alpha^2}{2}\frac{g^2_Z\gamma}{g^2_{2\text{ph}}}.
$$
This rate needs to be much faster than the single photon loss rate of the cat qubit modes to ensure a low phase-flip probability below the threshold of the biased-noise-tailored codes.
Finally, we note that, in principle, it is possible to increase this measurement rate by increasing the parameter $\nu$, leading to squeezed pointer states. Indeed, while the signal corresponding to the $X$ quadrature of this outgoing field is amplified by a factor $1/(1-\nu^2)$, the standard deviation of this quadrature only increases as $1/\sqrt{1-\nu^2}$. We note, however, that this on-chip amplification of the multi-Z signal could lead to undesirable effects on the natural functioning of cat qubits through, for instance, higher order nonlinearities that are neglected in this analysis. We postpone a thorough analysis of such strategies to future work. 

\section{Numerics}\label{sec:performances}
In this section, we numerically analyze the performance of logical qubits encoded using the QND multi-$Z$ measurement primitive.
In a first subsection, we study the case of moderate noise bias ($\eta = 10^2$ to $10^4$), where both bit-flip and phase-flip errors must be actively corrected using an asymmetric XZZX surface code.
In a second subsection, we consider the regime of very large noise bias ($\eta = 10^4$ to $10^{10}$), where the recent study of~\cite{shanahan2026} indicates that the optimal solution is to only focus on phase-flip correction for instance using a repetition code, and take care of residual bit-flips through concatenation with high-rate bit-flip codes.

\subsection{Moderate noise bias}\label{ssec:moderate_bias}

We consider bias values $\eta \in \{10^2, 10^3, 10^4\}$.
In this regime bit-flip errors occur at rate $\frac{p_z}{\eta}$, small but non-negligible,
so both $X$- and $Z$-type errors must be actively corrected. The recent study of~\cite{shanahan2026,Ruiz2026_Review} illustrates that the asymmetric XZZX code~\cite{bonilla2021xzzx} is a good choice for error correction with such moderate noise biases. It exploits the bias to achieve a large threshold for phase-flip errors while
maintaining a separate tunable distance for bit-flip protection.
Each XZZX stabilizer is measured with a single $M_{Z^4}$ and four
$\mathrm{CZ}$ gates as in Figure~\ref{fig:CX}(c). In this subsection, we analyze the performance of such an asymmetric XZZX encoding, implemented with QND multi-Z primitive under realistic circuit-level biased noise model.

\textit{Noise model.}
Each qubit is subject to a biased Pauli channel: after every elementary operation,
a phase-flip error may occur on each qubit with probability $p_z$ and a bit-flip type error with probabilities 
$p_x = p_y = p_z / (2\eta)$ such that $\eta = p_z/(p_x+p_y)$. {This biased Pauli channel is later referred to as $\texttt{PAULI\_1}(p_x/2,\;p_x/2,\;p_z)$}
The $M_{Z^4}$ assignment error probability is set to $\frac{p_z}{\eta}$, consistent with the
assumption that the MZ$^4$ assignment fidelity is limited by the physical bit-flip rate
(Section~\ref{sec:physical}); $M_X$ assignment error probability is $p_z$.
Syndrome extraction circuits are implemented in \texttt{stim}~\cite{gidney2021stim}. 
Logical errors are decoded by minimum-weight perfect matching via
\texttt{PyMatching}~\cite{higgott2022pymatching}, and statistics are collected with
\texttt{sinter}~\cite{gidney2021stim} using up to $10^9$ shots per task
and stopping after $10^3$ observed logical errors.

\begin{figure*}[t]
    \centering \includegraphics[width=\linewidth]{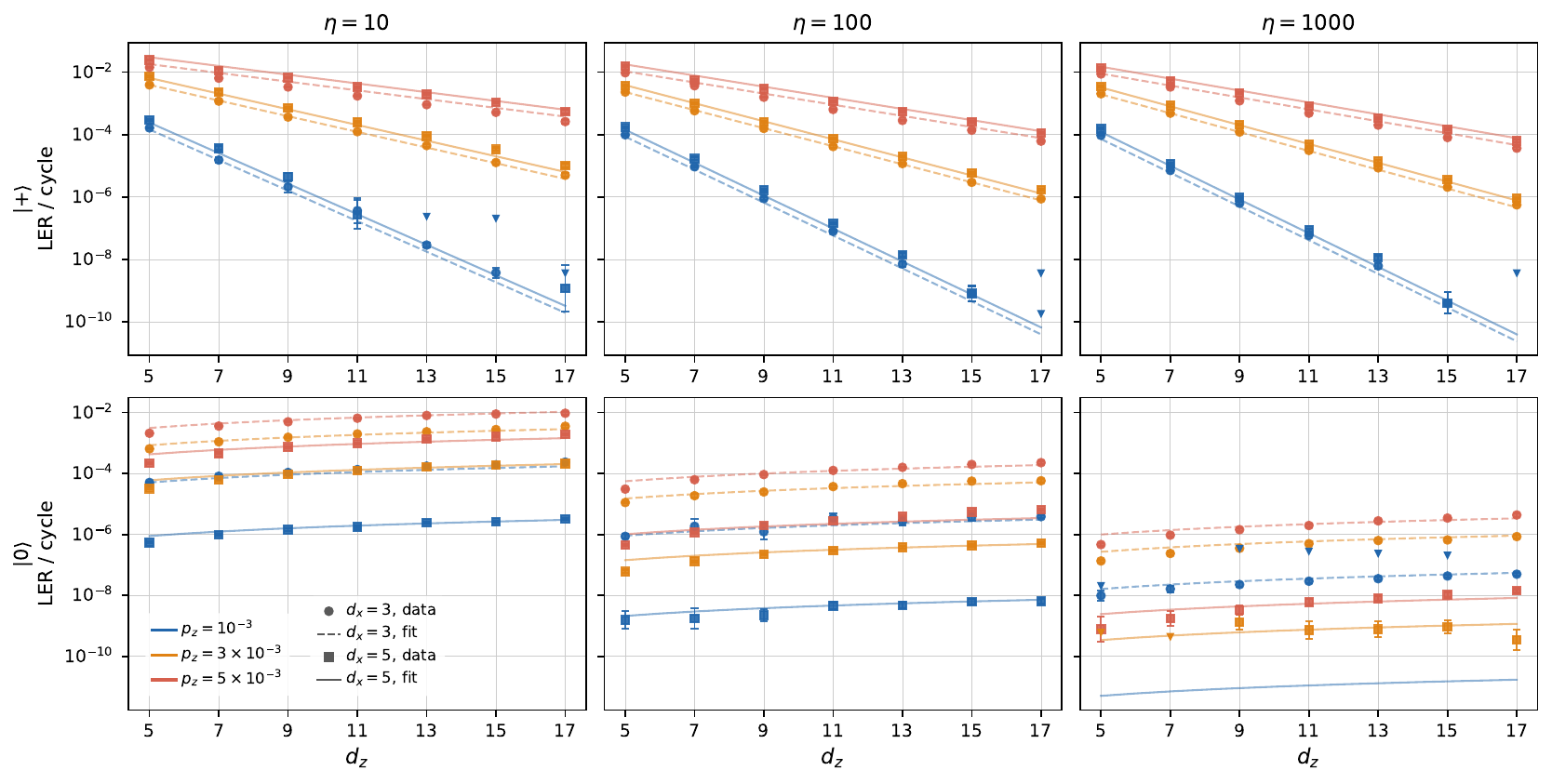}
    \caption{%
        Logical error rate per syndrome cycle of the XZZX surface code implemented
        with $M_{Z^4}$ measurements, as a function of $d_z$.
        \emph{Top row:} $\ket{+}$ memory (X-type logical).
        \emph{Bottom row:} $\ket{0}$ memory (Z-type logical).
        Columns correspond to bias $\eta \in \{10, 100, 10^3\}$.
        Colors indicate phase-flip rate $p_z$
        (blue: $10^{-3}$; orange: $3\!\times\!10^{-3}$; red: $5\!\times\!10^{-3}$);
        solid and dashed lines correspond to $d_x = 3$ and $d_x = 5$, respectively.
        Downward triangles ($\triangledown$) mark 95\% Poisson upper bounds for
        tasks with zero observed errors.
        Thin lines show the power-law ansätze of  Eqs.~\eqref{eq:ansatz_plus}--\eqref{eq:pth_eta}.
    }
    \label{fig:xzzx}
\end{figure*}

\textit{Results.}
Figure~\ref{fig:xzzx} shows the logical error rate (LER) per syndrome cycle for both the
$\ket{+}$ (X-type logical) and $\ket{0}$ (Z-type logical) memory experiments,
as a function of $d_z$ for $d_x \in \{3, 5\}$,
phase-flip rates $p_z \in \{10^{-3},\, 3\!\times\!10^{-3},\, 5\!\times\!10^{-3}\}$,
and bias values $\eta \in \{10, 100, 10^3\}$.
For $\eta \geq 100$ the LER decreases exponentially with $d_z$ across all tested
parameter regimes, demonstrating clear sub-threshold behavior.
At $\eta = 10^3$ and $p_z = 10^{-3}$, the $\ket{+}$ LER falls below $10^{-7}$ per
syndrome cycle already at $(d_x, d_z) = (3, 11)$.
At $\eta = 10$ and $p_z = 5\!\times\!10^{-3}$ the $\ket{+}$ curves show reduced
error suppression, consistent with the expected reduction in effective threshold at lower bias.

\textit{Sub-threshold scaling ansatz.}
At sub-threshold noise rates, the logical error rate follows a power-law scaling.
For the $\ket{+}$ memory,
\begin{equation}
    p_L^{(+)} = A\, d_x \left(\frac{p_z}{p_{z,\mathrm{th}}(\eta)}\right)^{(d_z+1)/2},
    \label{eq:ansatz_plus}
\end{equation}
where $A$ is a dimensionless prefactor and $p_{z,\mathrm{th}}(\eta)$ is an
$\eta$-dependent effective threshold that saturates to
$p_\infty \approx 1.25\%$ at large bias.
For the $\ket{0}$ memory,
\begin{equation}
    p_L^{(0)} = A'\, d_z
    \left(\frac{p_z}{\eta^\beta\, p_{x,\mathrm{th}}}\right)^{\alpha (d_x+1)/2},
    \label{eq:ansatz_zero}
\end{equation}
with $\alpha \approx 1.26$, $\beta \approx 0.68$, and
$p_{x,\mathrm{th}} \approx 5.3\!\times\!10^{-3}$.
The thin lines in Figure~\ref{fig:xzzx} show these ansätze overlaid on the data;
the detailed fitting procedure, including the parametrization of
$p_{z,\mathrm{th}}(\eta)$, is presented in Appendix~\ref{sec:fitting}.

\begin{figure}[ht]
    \centering
    \includegraphics[width=\linewidth]{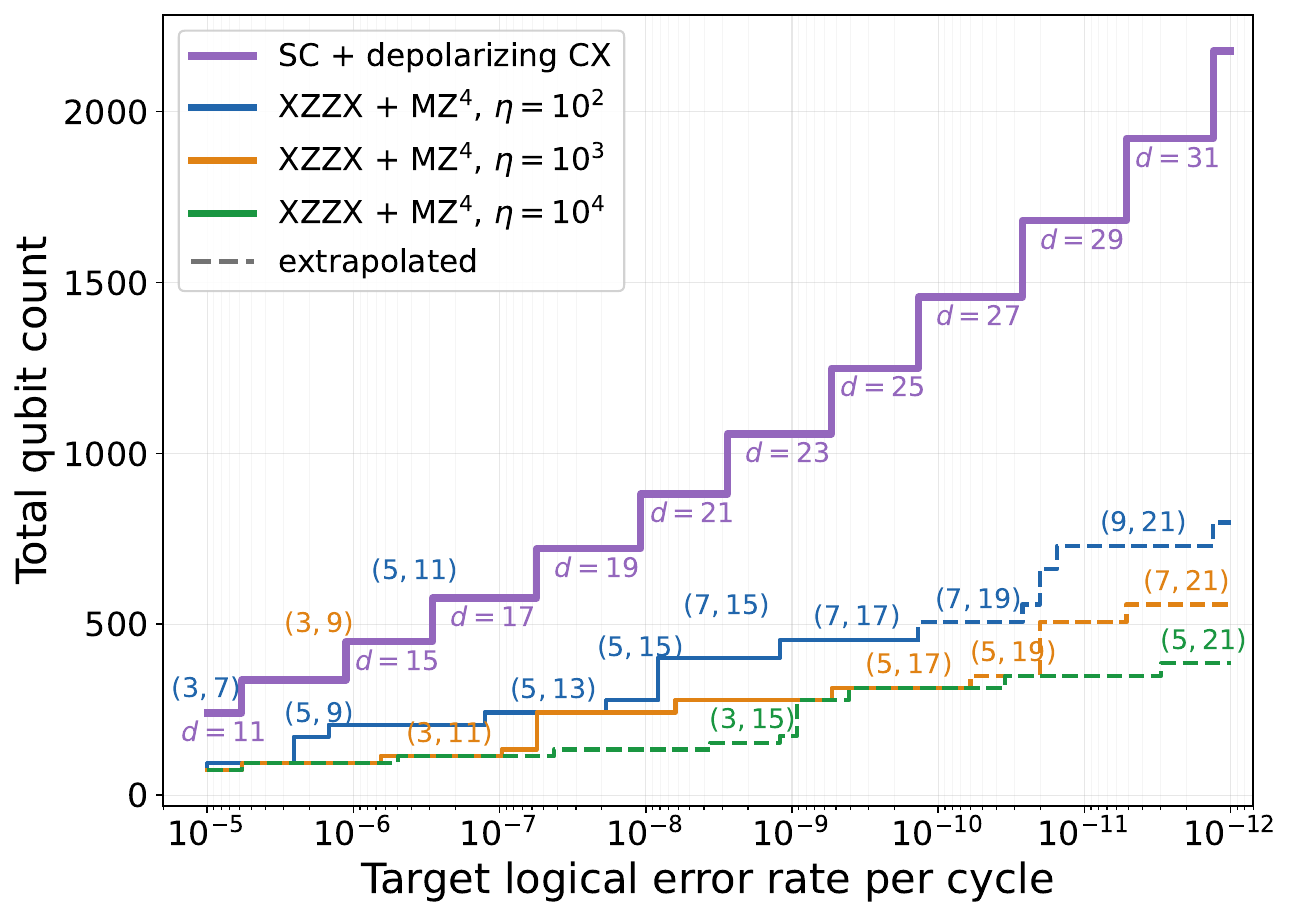}
    \caption{%
        Minimum qubit overhead to achieve a target logical error rate per cycle,
        at fixed $p_z = 10^{-3}$ and noise bias $\eta \in \{10^2, 10^3, 10^4\}$.
        \emph{Colored staircases:} XZZX$+$MZ$^4$ code; labels indicate the
        optimal $(d_x, d_z)$ at each step.
        \emph{Purple staircase:} rotated surface code with depolarizing CX
        (SC + depolarizing CX); labels show optimal distance $d$.
        Dashed portions indicate extrapolation beyond the fitted range
        ($\eta \leq 10^3$).
    }
    \label{fig:overhead}
\end{figure}

\textit{Overhead comparison.}
We now quantify the hardware efficiency of the XZZX$+$MZ$^4$ approach by comparing
its qubit overhead against a bias-unaware strategy at the same physical noise level.
The natural baseline is a standard rotated surface code (distance $d$,
$n_\mathrm{SC} = 2d^2 - 1$ qubits) using conventional depolarizing CNOT gates.
A CNOT applied to biased-noise qubits mixes all error channels,
yielding a total depolarizing rate
\begin{equation}
    p_\mathrm{cx} = 2p_z(1 + \eta^{-1}) \approx 2p_z.
    \label{eq:pcx}
\end{equation}
The depolarizing model for the two-qubit gate is motivated by the fact that
a conventional (non-bias-preserving) CNOT or CZ gate could depolarize the
dominant $Z$-type noise, so the output error channel is approximated by
a depolarizing channel whose total rate equals the sum of the input error rates
on both qubits.

All single-qubit operations (gates, resets, idle) retain the native
bias-preserving noise channel
$\texttt{PAULI\_1}(\frac{p_z}{2\eta},\;\frac{p_z}{2\eta},\;p_z)$.
Using fitted thresholds $p_\mathrm{th}^{(X)} \approx 0.46\%$
and $p_\mathrm{th}^{(Z)} \approx 0.68\%$ from our own numerical
simulations (see Appendix~\ref{sec:sc_simulations} for details),
we estimate the rotated-code overhead.
For the XZZX code with parameters $(d_x, d_z)$, the total physical footprint under
the alternating MZ$^4$ schedule is
\begin{equation}
    n_\mathrm{XZZX}(d_x, d_z) = 4\,d_x d_z + 2(d_x - d_z - 1).
    \label{eq:nxzzx}
\end{equation}
At leading order the footprint scales as $4\,d_x d_z$, roughly twice the
$2d_x d_z - 1$ qubits of a standard rotated XZZX surface code. This doubling
arises because the code is laid out on the non-rotated
lattice rather than the rotated one (see Appendix~\ref{sec:sc_simulations} for
the detailed layout and scheduling). The
subleading correction $2(d_x - d_z - 1)$ accounts for additional boundary
qubits that are reset every round.

\begin{figure}[t]
    \centering
    \raisebox{7em}{\resizebox{.8\linewidth}{!}{\begin{quantikz}[row sep=0.5em]
			\lstick{$\ket{+}$} & \gate[4, disable auto height][1][1]{\begin{matrix}
					M_{Z^4}\\ 	\downarrow\\s_1
			\end{matrix}}&&\gate{Z^{m_1}}&&&\gate{X^{s_1+s_2+s_3+s_4}}& \\
			\lstick{$D_0$} &&\meterD{M_X}&\setwiretype{n} \midstick{\hspace{-2.3em}$=m_1\quad\ket{+}$}&\setwiretype{q}&&&\\
			\lstick{$\ket{+}$} &&&\gate{Z^{m_2}}&\gate[4, disable auto height][1][1]{\begin{matrix}
					M_{Z^4}\\ 	\downarrow\\s_2
			\end{matrix}}\setwiretype{q}& \meterD{M_X}&\setwiretype{n}\midstick{\hspace{-5.1em}$= m_5\quad \ket{+}$}&\setwiretype{q}\\
			\lstick{$D_1$} &&\meterD{M_X} &\setwiretype{n}\midstick{\hspace{-2.3em}$=m_2\quad\ket{+}$}&\setwiretype{q}&&\gate{Z^{m_5}}&\\
			\lstick{$\ket{+}$} & \gate[4, disable auto height][1][1]{\begin{matrix}
					M_{Z^4}\\ 	\downarrow\\s_3
			\end{matrix}}&&\gate{Z^{m_3}}&& \meterD{M_X}&\setwiretype{n}\midstick{\hspace{-5.1em}$= m_6\quad \ket{+}$}&\setwiretype{q}\\
			\lstick{$D_2$} &&\meterD{M_X} &\setwiretype{n}\midstick{\hspace{-2.3em}$=m_3\quad\ket{+}$}&\setwiretype{q}&&\gate{Z^{m_6}}&\\
			\lstick{$\ket{+}$} &&&\gate{Z^{m_4}}&\gate[4, disable auto height][1][1]{\begin{matrix}
					M_{Z^4}\\ 	\downarrow\\s_4
			\end{matrix}}\setwiretype{q}& \meterD{M_X}&\setwiretype{n}\midstick{\hspace{-5.1em}$= m_7\quad \ket{+}$}&\setwiretype{q}\\
			\lstick{$D_3$} &&\meterD{M_X} &\setwiretype{n}\midstick{\hspace{-2.3em}$=m_4\quad\ket{+}$}&\setwiretype{q} &&\gate{Z^{m_7}}&\\
			\lstick{$\ket{+}$} &&&&&&\gate{Z^{m_8}}& \\
			\lstick{$D_4$} &&&&& \meterD{M_X}&\setwiretype{n}\midstick{\hspace{-5.1em}$= m_8\quad \ket{+}$}&\setwiretype{q}\\
		\end{quantikz}}}
    \\
    \includegraphics[width=0.7\linewidth]{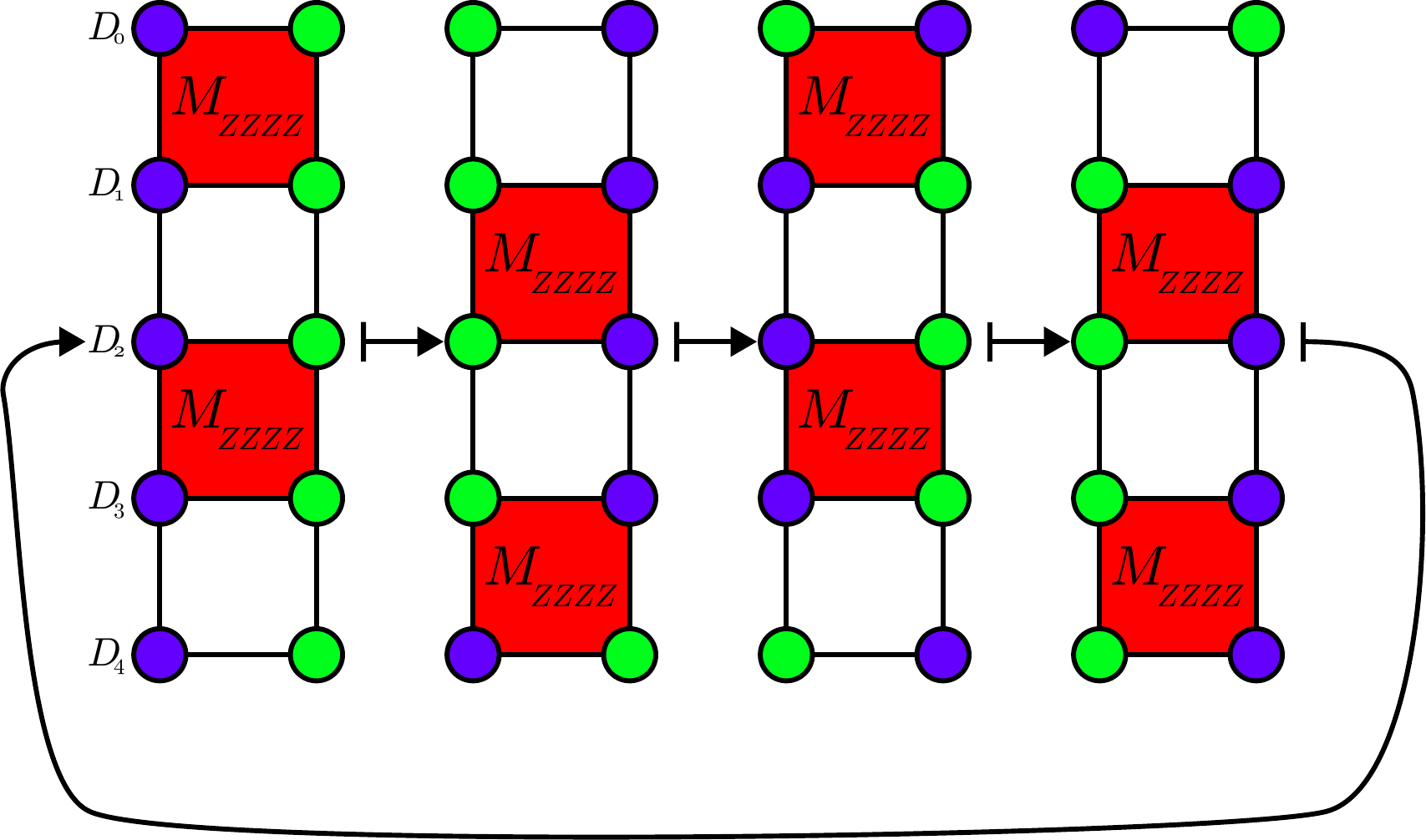}
    \caption{%
        Circuit and layout for the repetition code syndrome extraction using
        $M_{Z^4}$ and $2d$ qubits (alternating schedule, $d=5$).
        \emph{Left:} the circuit for two consecutive steps, yielding a full
        syndrome $\sigma = (m_1\oplus m_2,\, m_3\oplus m_4,\, m_5\oplus m_6,\, m_7\oplus m_8)$.
        \emph{Right:} 2D layout schematic.
        Purple vertices: data qubit positions at the start of each step.
        Green vertices: ancilla qubits initialized in $\ket{+}$.
        Red squares: $M_{Z^4}$ operations.
        After each step, data qubits involved in an $M_{Z^4}$ are
        measured in $X$ and teleported to the ancilla positions.
        Two steps extract a full syndrome; four steps return data to their
        initial positions.
    }
    \label{fig:mz4_alternated_circuit}
\end{figure}

For fixed $p_z = 10^{-3}$ and a target LER $\ell$, we minimise $n_\mathrm{XZZX}$
over integer $(d_x, d_z)$ subject to $p_L^{(+)} + p_L^{(0)} \leq \ell$.
The optimum lies near the balanced point $p_L^{(+)} \approx p_L^{(0)}$; as $\eta$
grows, the optimal aspect ratio $d_z/d_x$ increases, yielding an elongated code
block that concentrates resources on the dominant phase-flip channel.

Figure~\ref{fig:overhead} shows the resulting minimum qubit count as a function of
target LER for $\eta \in \{10^2, 10^3, 10^4\}$.
The purple staircase (SC + depolarizing CX) is nearly independent of $\eta$, while the
XZZX staircases fall steeply with increasing bias.
Quantitatively:
\begin{itemize}
    \item At $\eta = 10^2$, the XZZX code provides an advantage of
          $2.3$--$2.4\times$ across the LER range $10^{-7}$--$10^{-11}$.
    \item At $\eta = 10^3$, the advantage grows to $3.3$--$5.1\times$.
          For example, reaching LER $= 10^{-9}$ requires 278~qubits with
          $(d_x, d_z) = (5, 15)$, versus 1057~qubits ($d = 23$) for the
          depolarizing surface code.
    \item At $\eta = 10^4$ (extrapolated beyond the simulation range), the gain
          reaches $5.1\times$ at LER $= 10^{-7}$ and $4.8\times$ at $10^{-11}$.
\end{itemize}
Despite the $2\times$ qubit inflation inherent to the MZ$^4$ schedule, the
bias-preserving nature of the approach more than compensates: by concentrating
resources along the dominant phase-flip direction ($d_z \gg d_x$), the
XZZX$+$MZ$^4$ code achieves the same target logical error rate with
significantly fewer total qubits than a bias-unaware surface code, and the
advantage grows rapidly with increasing noise bias.

\subsection{Very large noise bias}\label{ssec:large_bias}

We now consider the regime of large noise biases ($\eta \approx 10^{4}$--$10^{10}$). As discussed in~\cite{shanahan2026}, in this case, an optimal encoding strategy is to neglect the rare bit-flips and focus on correcting phase-flip errors, for instance, with a repetition code. The residual bit-flips can then be corrected through a concatenation with a high-rate bit-flip code at very low cost. In this subsection, we analyze through  numerical simulations the performance of such a repetition code under circuit-level biased noise model. In particular, we provide a thorough comparison between the case where such an encoding is based on an $M_{Z^{4}}$ primitive, and the case where it relies on a bias-preserving CNOT primitive.

\textit{Syndrome extraction schedule.}
As shown in Figure~\ref{fig:mz4_alternated_circuit}, the repetition code
X-stabilizers are extracted via $M_{Z^4}$ measurements in an
alternating schedule: at each step, non-overlapping groups of four qubits
(two data, two ancilla) undergo a joint $Z^4$ measurement, after which
the involved data qubits  are measured in the $X$ basis and teleported to the
ancilla positions.
Two consecutive steps constitute a full syndrome extraction round (all $d-1$
stabilizers measured once).
The alternating schedule uses $2d$~qubits for a distance-$d$ repetition code
(Figure~\ref{fig:mz4_alternated_circuit}). Note that a different scheduling, called the simultaneous schedule, is possible and is presented Appendix~\ref{sec:alt_vs_sim}.

\begin{figure*}[t]
    \centering
    \includegraphics[width=0.85\textwidth]{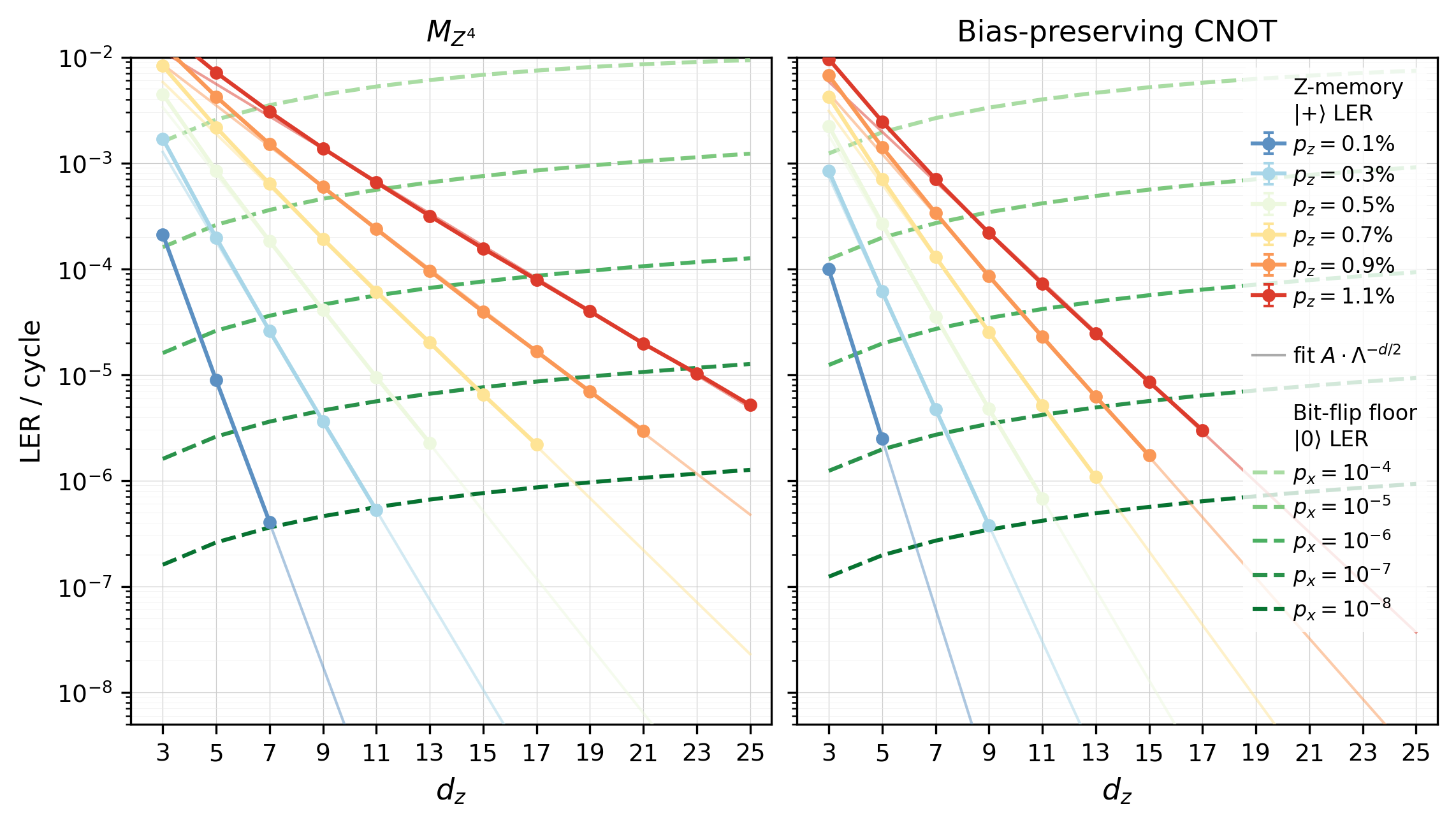}
    \caption{%
        Phase-flip logical error rate per cycle for the repetition code,
        comparing the $M_{Z^4}$ alternated scheme (left) and
        the bias-preserving CNOT scheme (right).
        Solid lines show the $\ket{+}$-memory (phase-flip) LER per cycle
        as a function of distance $d$.
        Dashed lines: unprotected bit-flip floor (from DEM product formula).
        Thin solid lines: exponential fits $A(p_z)\,\Lambda^{-d/2}$.
    }
    \label{fig:repcode_comparison}
\end{figure*}

\textit{Noise model.}
We compare two syndrome extraction schemes for the repetition code under
physically motivated circuit-level noise.
In both schemes, idle qubits experience a biased Pauli channel
$\texttt{PAULI\_1}(\frac{p_x}{2},\frac{p_x}{2},p_z)$ at every time step.

For the \emph{MZ$^4$ alternated scheme} ($2d$ qubits, 4 time steps per round):
the $M_{Z^4}$ measurement has an assignment error probability $p_x$
(limited by the physical bit-flip rate);
the $M_X$ teleportation measurements flip with probability $p_z$;
and the idle noise channel is applied to \emph{all} qubits after each $M_{Z^4}$,
and to non-measured qubits after each $M_X$.

For the \emph{bias-preserving CNOT scheme} ($2d-1$ qubits, 3 time steps per round):
each CX gate is followed by a two-qubit Pauli channel with dominant $Z$-type
errors ($IZ$, $ZI$, $ZZ$ each at rate $\frac{2p_z}{3}$) and rare bit-flip errors
($IX$, $XI$, $XX$ each at rate $\frac{2p_x}{3}$);
the ancilla $M_X$ readout flips with probability $p_z$;
and the idle noise is applied only to the data qubit not participating in the
current CX layer, and to all data qubits during the ancilla measurement operation.

Table~\ref{tab:repcode_noise} summarizes the noise channels for both schemes.
The key difference is that in the MZ$^4$ scheme, data qubits are
never subject to a conditional unitary and accumulate only idle phase-flip
errors, whereas the CNOT gate introduces correlated $ZZ$ errors (phase kickback)
on top of the independent $Z$ errors on each qubit.

\begin{table}[h!]
\centering
\caption{Circuit-level noise channels for the two repetition code schemes.}
\label{tab:repcode_noise}
\begin{tabular}{|l|c|c|}
\hline
\textbf{Operation} & \textbf{MZ$^4$ alt.} & \textbf{BP-CNOT} \\
\hline
Idle & \multicolumn{2}{c|}{$\texttt{PAULI\_1}(\frac{p_x}{2},\frac{p_x}{2}, p_z)$} \\\hline
$M_{Z^4}$ assignment err. & $p_x$ & --- \\
$M_X$ assignment err. & $p_z$ & $p_z$ \\
CX (2-qubit) & --- & $IZ,ZI,ZZ: \frac{2p_z}{3}$ \\
 & & $IX,XI,XX: \frac{2p_x}{3}$ \\
\hline\hline
Qubits & $2d$ & $2d-1$ \\
Time steps / round & 4 & 3 \\
\hline
\end{tabular}
\end{table}

The simulations are performed in \texttt{stim}~\cite{gidney2021stim} for
$d \in \{3, 5, \ldots, 25\}$, $\mathrm{rounds} = d$, and
$p_z \in \{1, 3, 5, 7, 9, 11\} \times 10^{-3}$.
The bit-flip rate $p_x$ is set small enough at each $p_z$ {so} that the
bit-flip floor remains negligible relative to the phase-flip LER
at the simulated distances.
Statistics are collected up to $10^7$~shots or $1.5\times 10^4$~observed logical errors per task;
decoding is performed with minimum-weight perfect matching
(\texttt{PyMatching}~\cite{higgott2022pymatching}).

\begin{figure}[t]
    \centering
    \includegraphics[width=\columnwidth]{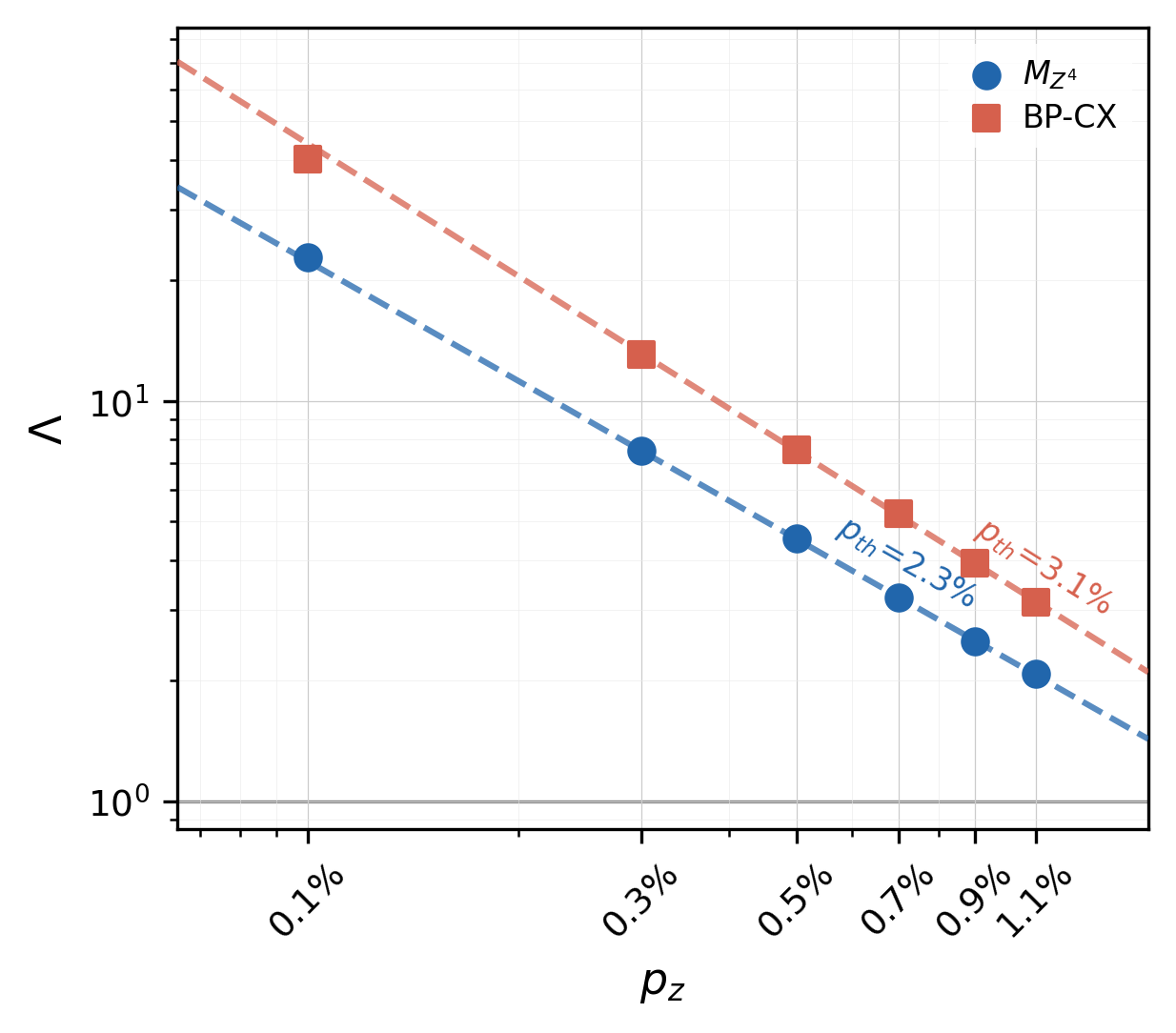}
    \caption{%
        Error suppression factor $\Lambda$ as a function of $p_z$ for the
        $M_{Z^4}$ alternated and bias-preserving CNOT repetition
        code schemes.
        $\Lambda$ is extracted from the exponential slope of LER vs.\ $d$
        [Eq.~\eqref{eq:repcode_ansatz}].
        Dashed lines: power-law fits; annotations indicate the
        effective thresholds $p_{z,\mathrm{th}}^{\mathrm{eff}}$ (where $\Lambda = 1$).
        The analytical estimates $p_{z,\mathrm{th}} \approx 2.5\%$ (MZ$^4$,
        $n_{\mathrm{eff}}=4$) and $\approx 3.3\%$ (BP-CNOT, $n_{\mathrm{eff}}=3$),
        derived from $p_{\mathrm{th}}^{\mathrm{pheno}} \approx 10\%$, are in good
        agreement with the fitted values ($2.3\%$ and $3.1\%$ respectively).
    }
    \label{fig:repcode_lambda}
\end{figure}

\textit{Results.}
Figure~\ref{fig:repcode_comparison} shows the phase-flip logical error rate (LER)
per syndrome cycle as a function of code distance $d$ for both schemes:
$M_{Z^4}$ alternated and bias-preserving CNOT.
In the sub-threshold regime, the LER follows an exponential decay
\begin{equation}
    p_L^{(Z)} \approx A(p_z)\,\Lambda(p_z)^{-d/2},
    \label{eq:repcode_ansatz}
\end{equation}
where $\Lambda(p_z) = p_{z,\mathrm{th}}^{\mathrm{eff}}/p_z$ controls the error
suppression rate per unit distance and $A(p_z)$ is a scheme-dependent prefactor
(in principle constant, but the simulations show a mild $p_z$-dependence; since
we do not extrapolate in $p_z$, this does not affect the overhead estimates).
The dashed lines show the unprotected bit-flip floor computed from the
detector error model~\cite{gidney2021stim}, confirming that it remains orders of magnitude below
the phase-flip LER across all simulated parameters.

\textit{Error suppression and effective thresholds.}
Figure~\ref{fig:repcode_lambda} shows the extracted error suppression factor
$\Lambda(p_z)$ for each scheme.
Linear fits in the log-log plot yield effective thresholds
$p_{z,\mathrm{th}}^{\mathrm{eff}}$ (the $p_z$ at which $\Lambda = 1$).
These thresholds can be understood from the well-known phenomenological
threshold of the repetition code, $p_{\mathrm{th}}^{\mathrm{pheno}} \approx 10\%$~\cite{dennis2002topological}.
At the circuit level, the effective threshold is reduced by the number of
error mechanisms per syndrome cycle: $p_{z,\mathrm{th}}^{\mathrm{eff}} \approx
p_{\mathrm{th}}^{\mathrm{pheno}} / n_{\mathrm{eff}}$, where $n_{\mathrm{eff}}$
counts the effective number of independent $Z$-error opportunities per data qubit
per round.
For the MZ$^4$ alternating schedule, an interior data qubit accumulates
$Z$ errors from 4 time steps (2 $M_{Z^4}$ steps, 2 $M_X$ steps),
giving $n_{\mathrm{eff}} \approx 4$ and thus
$p_{z,\mathrm{th}}^{\mathrm{MZ4}} \approx 10\%/4 = 2.5\%$.
For the BP-CNOT scheme, each data qubit participates in 2 CX gates
plus one MRX readout per cycle---i.e.\ 3 error-injection points---giving
$n_{\mathrm{eff}} = 3$ and
$p_{z,\mathrm{th}}^{\mathrm{CNOT}} \approx 10\%/3 \approx 3.3\%$,
consistent with the numerically fitted value of $3.1\%$ \footnote{A finer counting---$IZ$ and $ZZ$ each at $2p_z/3$ per CX gate, giving a marginal $Z$-rate of $4p_z/3$ per layer and $n_{\mathrm{eff}}\approx 11/3$---predicts $\approx 2.7\%$ and undershoots the numerical threshold.
The reason is that the correlated $ZZ$ term creates simultaneous ancilla and data $Z$ errors that produce a characteristic detector pattern (adjacent syndrome defects at the same time step), which the MWPM decoder handles more easily than two independent $Z$ errors of the same total weight.
The ``3 events'' counting, which treats each CX gate as a single error-injection point regardless of internal correlations, captures the decoder-effective noise more faithfully.
Importantly, the threshold ratio $3.1\%/2.3\% \approx 4/3$ directly reflects the ratio of error-injection points per cycle ($4$ for MZ$^4$ vs.\ $3$ for BP-CNOT).}.

\begin{figure}[t]
    \centering
    \includegraphics[width=0.5\textwidth]{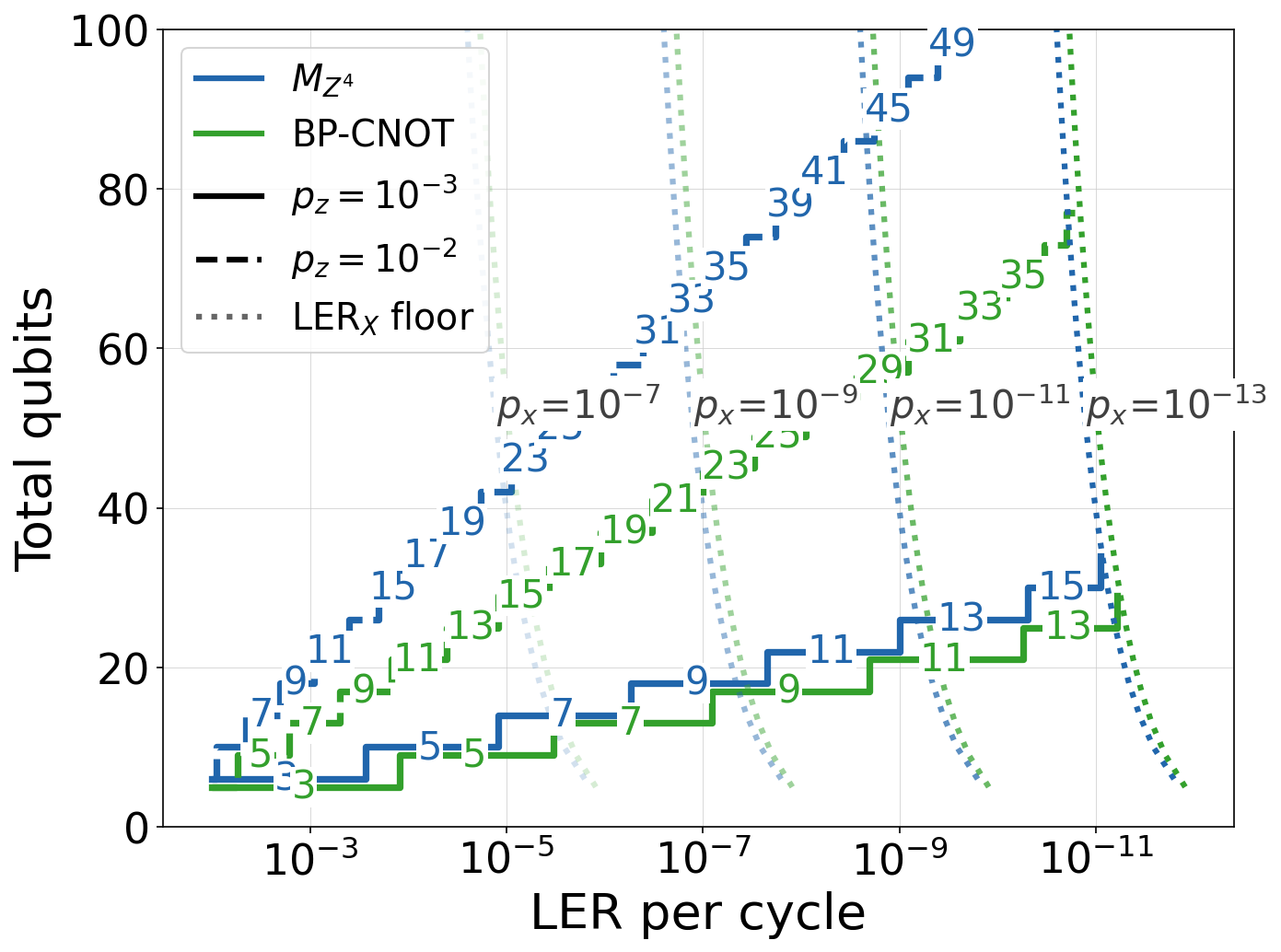}
    \caption{%
        Qubit overhead comparison for the repetition code at very large noise bias.
        Solid lines: $p_z = 10^{-3}$; dashed: $p_z = 10^{-2}$.
        Opacity encodes the bit-flip rate $p_x$ (faint: $10^{-7}$, opaque: $10^{-13}$).
        For each target logical error rate per cycle (x-axis), the minimum number of
        physical qubits is shown for: $M_{Z^4}$ alternated ($2d$ qubits,
        blue) and bias-preserving CNOT ($2d-1$ qubits, green).
        Dotted diagonal lines indicate the bit-flip floor: below this LER, increasing
        $d$ no longer helps.
    }
    \label{fig:repcode_overhead}
\end{figure}

\textit{Overhead comparison.}
Finally, we compare the total qubit overhead required to reach a target
logical error rate for both schemes at two physical phase-flip rates
$p_z \in \{10^{-3}, 10^{-2}\}$ and bit-flip rates
$p_x \in \{10^{-7}, 10^{-9}, 10^{-11}, 10^{-13}\}$.
At finite $p_x$, the bit-flip floor limits the minimum achievable
LER; its contribution is computed from the detector error model and
extrapolated linearly with $d$.
The total LER is $p_L = p_L^{(Z)} + p_L^{(X)}$, and the overhead is
minimized over $d$ at each target $p_L$.

Figure~\ref{fig:repcode_overhead} presents the results.
The BP-CNOT scheme achieves a slightly better overhead across
target LER values, owing to its slightly higher phase-flip error threshold.
Importantly, the MZ$^4$ scheme achieves comparable performance without
requiring a bias-preserving CNOT gate, which as discussed in the
introduction remains experimentally challenging.

\section{Conclusion}\label{sec:conclusion}
The noise bias present in some quantum computing platforms can be combined with tailored error correcting codes to reduce significantly the hardware overhead required for fault-tolerance.
It has been generally admitted that this relied on the availability of a CNOT operation which does not convert frequent phase-flips to rare bit-flips.
Such a bias-preserving CNOT operation is forbidden for qubits defined in a two-dimensional Hilbert space~\cite{guillaud2019repetition}, strongly restricting the class of biased-noise qubits considered to be able to benefit from a tailored error correction.

In this paper, we demonstrate that an alternative approach exists for operating biased-noise qubits of all types and, therefore, they all can benefit from such tailored error correction. This relies on the implementation of ultra high fidelity multi-qubit quantum non-demolition measurements of Pauli $Z$ operators. We provide theoretical proposals for physical implementations of these primitive operations in two settings: nuclear spins coupled to electron spins via hyperfine interaction, and dissipative cat qubits implemented with superconducting circuits. We propose optimized compiling of tailored error correction logical circuits based on such primitive operations and numerically evaluate the hardware resource savings that are provided by the noise bias. These numerical simulations indicate that, with these  primitive operations, it is possible to ensure a significant gain in the hardware overhead of error correction, similar to the case where we have access to an efficient bias-preserving CNOT. 

Note that this primitive also opens the route to the implementation of other recent proposals for hardware-efficient fault-tolerant computation. In a recent work~\cite{Ruiz-LDPC-2025}, we demonstrated that in the case of large noise biases, one can replace the phase-flip repetition code with a 2D classical LDPC code, leading to a high-rate encoding of logical qubits and drastically reducing the overhead of error correction. Furthermore, it is possible to  operate the logical qubits in a code block through lattice surgery techniques in a bilayer architecture. All these protocols rely on the implementation of weight-4 $X$ parity checks that are ideally realized with bias-preserving CNOTs between data qubits and a  measurement ancilla qubit. In Appendix~\ref{sec:ldpc_parity}, we demonstrate that it is also possible to compile such parity checks with a multi-qubit QND Pauli $Z$ measurement primitive. The same weight-4 parity checks are also used in another recent work~\cite{Ruiz-npj-2026} to achieve very low cost magic state distillation, even in the case of moderate noise bias. Based on these constructions, with an elementary set of bias-preserving operations $\{\mathcal{P}_{\ket{0}},\mathcal{P}_{\ket{+}},\mathcal{M}_X,\mathcal{M}^{\text{QND}}_{Z^3}\}$, and with the addition of a bias non-preserving operation $X^{\pm 1/4}$, we  obtain a universal set of hardware-efficient fault-tolerant operations at the logical level.

\section*{Acknowledgments}
The authors are grateful to Patrice Bertet, Emmanuel Flurin, Ronan Gautier, and Raphaël Lescanne for many enlightening discussions. We particularly thank Patrice Bertet for his comments on an early draft of this manuscript. 
This work was supported by the Plan France 2030 under
project ANR-22-PETQ-0006.
\bibliography{references}

@article{Hensen_Nat_Nano,
	author = {Hensen, Bas and Wei Huang, Wister and Yang, Chih-Hwan and Wai Chan, Kok and Yoneda, Jun and Tanttu, Tuomo and Hudson, Fay E. and Laucht, Arne and Itoh, Kohei M. and Ladd, Thaddeus D. and Morello, Andrea and Dzurak, Andrew S.},
	date = {2020/01/01},
	doi = {10.1038/s41565-019-0587-7},
	id = {Hensen2020},
	isbn = {1748-3395},
	journal = {Nature Nanotechnology},
	number = {1},
	pages = {13--17},
	title = {A silicon quantum-dot-coupled nuclear spin qubit},
	url = {https://doi.org/10.1038/s41565-019-0587-7},
	volume = {15},
	year = {2020}}

@article{Steinacker_PRL,
  title = {Coupling a ${}^{73}$Ge nuclear spin to an electrostatically defined quantum dot in silicon},
  author = {Steinacker, Paul and Goenka, Gauri and Su, Rocky Yue and Tanttu, Tuomo and Lim, Wee Han and Serrano, Santiago and Botzem, Tim and Cifuentes, Jesus D. and Lim, Shao Qi and McCallum, Jeffrey C. and Johnson, Brett C. and Hudson, Fay E. and Chan, Kok Wai and Escott, Christopher C. and Saraiva, Andre and Yang, Chih Hwan and Mourik, Vincent and Morello, Andrea and Dzurak, Andrew S. and Laucht, Arne},
  journal = {Phys. Rev. Lett.},
  pages = {},
  year = {2026},
  month = {May},
  publisher = {American Physical Society},
  doi = {10.1103/49gt-msw9},
  url = {https://link.aps.org/doi/10.1103/49gt-msw9}
}

@article{acharya2024surfacecode,
  author  = {Rajeev Acharya and Laleh Aghababaie-Beni and Igor Aleiner and others},
  title   = {Quantum error correction below the surface code threshold},
  journal = {Nature},
  volume  = {638},
  pages   = {920--926},
  year    = {2025},
  doi     = {10.1038/s41586-024-08449-y}
}

@article{aliferis2008biased,
  author  = {Panos Aliferis and John Preskill},
  title   = {Fault-tolerant quantum computation against biased noise},
  journal = {Phys. Rev. A},
  volume  = {78},
  pages   = {052331},
  year    = {2008},
  doi     = {10.1103/PhysRevA.78.052331}
}

@article{AWS-Blueprint-2022,
  title = {Building a Fault-Tolerant Quantum Computer Using Concatenated Cat Codes},
  author = {Chamberland, Christopher and Noh, Kyungjoo and Arrangoiz-Arriola, Patricio and Campbell, Earl T. and Hann, Connor T. and Iverson, Joseph and Putterman, Harald and Bohdanowicz, Thomas C. and Flammia, Steven T. and Keller, Andrew and Refael, Gil and Preskill, John and Jiang, Liang and Safavi-Naeini, Amir H. and Painter, Oskar and Brand\~ao, Fernando G.S.L.},
  journal = {PRX Quantum},
  volume = {3},
  issue = {1},
  pages = {010329},
  numpages = {117},
  year = {2022},
  month = {Feb},
  publisher = {American Physical Society},
  doi = {10.1103/PRXQuantum.3.010329},
  url = {https://link.aps.org/doi/10.1103/PRXQuantum.3.010329}
}

@article{bluvstein2023logical,
  author  = {Dolev Bluvstein and Simon J. Evered and Alexandra A. Geim and others},
  title   = {Logical quantum processor based on reconfigurable atom arrays},
  journal = {Nature},
  volume  = {626},
  pages   = {58--65},
  year    = {2023},
  doi     = {10.1038/s41586-023-06927-3}
}

@article{bonilla2021xzzx,
  title={The XZZX surface code},
  author={Bonilla Ataides, J Pablo and Tuckett, David K and Bartlett, Stephen D and Flammia, Steven T and Brown, Benjamin J},
  journal={Nature communications},
  volume={12},
  number={1},
  pages={2172},
  year={2021},
  publisher={Nature Publishing Group UK London}
}

@article{brooks-preskill-2013,
  title = {Fault-tolerant quantum computation with asymmetric Bacon-Shor codes},
  author = {Brooks, Peter and Preskill, John},
  journal = {Phys. Rev. A},
  volume = {87},
  issue = {3},
  pages = {032310},
  numpages = {16},
  year = {2013},
  month = {Mar},
  publisher = {American Physical Society},
  doi = {10.1103/PhysRevA.87.032310},
  url = {https://link.aps.org/doi/10.1103/PhysRevA.87.032310}
}

@article{campbell2017roads,
  author  = {Earl T. Campbell and Barbara M. Terhal and Christophe Vuillot},
  title   = {Roads towards fault-tolerant universal quantum computation},
  journal = {Nature},
  volume  = {549},
  pages   = {172--179},
  year    = {2017},
  doi     = {10.1038/nature23460}
}

@article{Claes-npj-2023,
	author = {Claes, Jahan and Bourassa, J. Eli and Puri, Shruti},
	doi = {10.1038/s41534-023-00677-w},
	journal = {npj Quantum Information},
	number = {1},
	pages = {9},
	title = {Tailored cluster states with high threshold under biased noise},
	url = {https://doi.org/10.1038/s41534-023-00677-w},
	volume = {9},
	year = {2023},
}

@INPROCEEDINGS{Cortinas-APS-2021,
       author = {{Corti{\~n}as}, Rodrigo and {Frattini}, Nicholas and {Puri}, Shruti and {Duke}, Owen and {Lei}, Chan U. and {Girvin}, Steven and {Devoret}, Michel},
        title = "{Toward a topological CNOT between two Kerr-cat qubits: part 1/2}",
    booktitle = {APS March Meeting Abstracts},
         year = 2021,
       series = {APS Meeting Abstracts},
       volume = {2021},
        month = jan,
          eid = {L33.006},
        pages = {L33.006},
}

@article{dennis2002topological,
  title={Topological quantum memory},
  author={Dennis, Eric and Kitaev, Alexei and Landahl, Andrew and Preskill, John},
  journal={Journal of Mathematical Physics},
  volume={43},
  number={9},
  pages={4452--4505},
  year={2002},
  publisher={AIP Publishing},
  doi={10.1063/1.1499754}
}

@misc{eisert2025,
      title={Mind the gaps: The fraught road to quantum advantage}, 
      author={Jens Eisert and John Preskill},
      year={2025},
      eprint={2510.19928},
      archivePrefix={arXiv},
      primaryClass={quant-ph},
      url={https://arxiv.org/abs/2510.19928}, 
}

@INPROCEEDINGS{Essig-APS-2023,
       author = {{Essig}, Antoine and {Bourdaud}, Nicolas and {Cohen}, Joachim and {Cottet}, Nathanael and {Devanz}, Louise and {Fevrier}, Pierre and {Gras}, Antoine and {Guillaud}, J{\'e}r{\'e}mie and {G{\"u}m{\"u}s}, Efe and {Hall{\'e}n}, Mattis and {Hasanuzzaman Kamrul}, Venus and {Jezouin}, Sebastion and {Lescanne}, Raphael and {Magnard}, Paul and {Roul}, Julien and {Pankratova}, Natalia and {Rautschke}, Felix and {Peronnin}, Theau and {Roverc'h}, Erwan and {Stevens}, Jeremy and {Polis}, Stephane and {Ville}, Jean-Loup and {Wan-Fat}, Pierre and {Rousseau}, R{\'e}mi and {Alice} and {Bob Team}},
        title = "{Towards a bias-preserving CNOT gate between stabilized cat qubits (Part 1)}",
    booktitle = {APS March Meeting Abstracts},
         year = 2023,
       series = {APS Meeting Abstracts},
       volume = {2023},
        month = jan,
          eid = {K75.008},
        pages = {K75.008},
}

@article{fowler2012surface,
  title={Surface codes: Towards practical large-scale quantum computation},
  author={Fowler, Austin G. and Mariantoni, Matteo and Martinis, John M. and Cleland, Andrew N.},
  journal={Physical Review A},
  volume={86},
  number={3},
  pages={032324},
  year={2012},
  publisher={American Physical Society},
  doi={10.1103/PhysRevA.86.032324}
}

@article{Frattini-PRX-2024,
  title = {Observation of Pairwise Level Degeneracies and the Quantum Regime of the Arrhenius Law in a Double-Well Parametric Oscillator},
  author = {Frattini, Nicholas E. and Corti\~nas, Rodrigo G. and Venkatraman, Jayameenakshi and Xiao, Xu and Su, Qile and Lei, Chan U. and Chapman, Benjamin J. and Joshi, Vidul R. and Girvin, S. M. and Schoelkopf, Robert J. and Puri, Shruti and Devoret, Michel H.},
  journal = {Phys. Rev. X},
  volume = {14},
  issue = {3},
  pages = {031040},
  numpages = {28},
  year = {2024},
  month = {Sep},
  publisher = {American Physical Society},
  doi = {10.1103/PhysRevX.14.031040},
  url = {https://link.aps.org/doi/10.1103/PhysRevX.14.031040}
}

@article{gidney2021stim,
  title     = {Stim: a fast stabilizer circuit simulator},
  author    = {Gidney, Craig},
  journal   = {Quantum},
  volume    = {5},
  pages     = {497},
  year      = {2021},
  publisher = {Verein zur F{\"o}rderung des Open Access Publizierens in den Quantenwissenschaften},
  doi       = {10.22331/q-2021-07-06-497},
  url       = {https://doi.org/10.22331/q-2021-07-06-497}
}

@misc{Grimm-Arxiv-2025,
      title={Enhancing Kerr-Cat Qubit Coherence with Controlled Dissipation}, 
      author={Francesco Adinolfi and Daniel Z. Haxell and Alessandro Bruno and Laurent Michaud and Venus Hasanuzzaman Kamrul and Preeti Pandey and Alexander Grimm},
      year={2025},
      eprint={2511.01027},
      archivePrefix={arXiv},
      primaryClass={quant-ph},
      url={https://arxiv.org/abs/2511.01027}, 
}

@article{grimm2020kerrcat,
  author  = {Alexander Grimm and Nicolas E. Frattini and Steven Puri and S. Shankar and Michel H. Devoret and others},
  title   = {Stabilization and operation of a Kerr-cat qubit},
  journal = {Nature},
  volume  = {584},
  pages   = {205--209},
  year    = {2020},
  doi     = {10.1038/s41586-020-2587-z}
}

@article{Gross-PRApp-2024,
  title = {Hardware-efficient error-correcting codes for large nuclear spins},
  author = {Gross, Jonathan A. and Godfrin, Cl\'ement and Blais, Alexandre and Dupont-Ferrier, Eva},
  journal = {Phys. Rev. Appl.},
  volume = {22},
  issue = {1},
  pages = {014006},
  numpages = {10},
  year = {2024},
  month = {Jul},
  publisher = {American Physical Society},
  doi = {10.1103/PhysRevApplied.22.014006},
  url = {https://link.aps.org/doi/10.1103/PhysRevApplied.22.014006}
}

@article{guillaud2019repetition,
  author  = {J{\'e}r{\'e}mie Guillaud and Mazyar Mirrahimi},
  title   = {Repetition Cat Qubits for Fault-Tolerant Quantum Computation},
  journal = {Phys. Rev. X},
  volume  = {9},
  pages   = {041053},
  year    = {2019},
  doi     = {10.1103/PhysRevX.9.041053}
}

@article{higgott2022pymatching,
  title         = {PyMatching: A Python package for decoding quantum codes with minimum-weight perfect matching},
  author        = {Higgott, Oscar and Gidney, Craig},
  journal       = {Quantum},
  volume        = {6},
  pages         = {817},
  year          = {2022},
  publisher     = {Verein zur Förderung des Open Access Publizierens in den Quantenwissenschaften},
  doi           = {10.22331/q-2022-10-24-817},
  url           = {https://doi.org/10.22331/q-2022-10-24-817}
}

@article{Jiang-PRL-2008,
  title = {Coherence of an Optically Illuminated Single Nuclear Spin Qubit},
  author = {Jiang, L. and Dutt, M. V. Gurudev and Togan, E. and Childress, L. and Cappellaro, P. and Taylor, J. M. and Lukin, M. D.},
  journal = {Phys. Rev. Lett.},
  volume = {100},
  issue = {7},
  pages = {073001},
  numpages = {4},
  year = {2008},
  month = {Feb},
  publisher = {American Physical Society},
  doi = {10.1103/PhysRevLett.100.073001},
  url = {https://link.aps.org/doi/10.1103/PhysRevLett.100.073001}
}

@article{Kane-Nature-98,
	author = {Kane, B.  E. },
	doi = {10.1038/30156},
	journal = {Nature},
	number = {6681},
	pages = {133--137},
	title = {A silicon-based nuclear spin quantum computer},
	url = {https://doi.org/10.1038/30156},
	volume = {393},
	year = {1998},
}

@article{Kruckenhauser-PRL-2025,
  title = {Dark Spin-Cat States as Biased Qubits},
  author = {Kruckenhauser, Andreas and Yuan, Ming and Zheng, Han and Mamaev, Mikhail and Zeng, Pei and Mao, Xuanhui and Xu, Qian and Zache, Torsten V. and Jiang, Liang and van Bijnen, Rick and Zoller, Peter},
  journal = {Phys. Rev. Lett.},
  volume = {135},
  issue = {2},
  pages = {020601},
  numpages = {9},
  year = {2025},
  month = {Jul},
  publisher = {American Physical Society},
  doi = {10.1103/w9zh-jwsx},
  url = {https://link.aps.org/doi/10.1103/w9zh-jwsx}
}

@article{leghtas2015manifold,
  author  = {Zaki Leghtas and Steven Touzard and Ioan M. Pop and others},
  title   = {Confining the state of light to a quantum manifold by engineered two-photon loss},
  journal = {Science},
  volume  = {347},
  number  = {6224},
  pages   = {853--857},
  year    = {2015},
  doi     = {10.1126/science.aaa2085}
}

@article{lescanne2020exponential,
  author  = {Rapha{\"e}l Lescanne and Marius Villiers and Th{\'e}au Peronnin and Alain Sarlette and Matthieu Delbecq and Benjamin Huard and Takis Kontos and Mazyar Mirrahimi and Zaki Leghtas},
  title   = {Exponential suppression of bit-flips in a qubit encoded in an oscillator},
  journal = {Nature Physics},
  volume  = {16},
  pages   = {509--513},
  year    = {2020},
  doi     = {10.1038/s41567-020-0824-x}
}

@Article{LesHouches-2023,
	title={{Quantum computation with cat qubits}},
	author={Jérémie Guillaud and Joachim Cohen and Mazyar Mirrahimi},
	journal={SciPost Phys. Lect. Notes},
	pages={72},
	year={2023},
	publisher={SciPost},
	doi={10.21468/SciPostPhysLectNotes.72},
	url={https://scipost.org/10.21468/SciPostPhysLectNotes.72},
}

@article{Marquet-PRX-2024,
  title = {Autoparametric Resonance Extending the Bit-Flip Time of a Cat Qubit up to 0.3 s},
  author = {Marquet, A. and Essig, A. and Cohen, J. and Cottet, N. and Murani, A. and Albertinale, E. and Dupouy, S. and Bienfait, A. and Peronnin, T. and Jezouin, S. and Lescanne, R. and Huard, B.},
  journal = {Phys. Rev. X},
  volume = {14},
  issue = {2},
  pages = {021019},
  numpages = {30},
  year = {2024},
  month = {Apr},
  publisher = {American Physical Society},
  doi = {10.1103/PhysRevX.14.021019},
  url = {https://link.aps.org/doi/10.1103/PhysRevX.14.021019}
}

@article{Marvian-PRXQ-2024,
  title = {Fault-Tolerant Quantum Computation Using Large Spin-Cat Codes},
  author = {Omanakuttan, Sivaprasad and Buchemmavari, Vikas and Gross, Jonathan A. and Deutsch, Ivan H. and Marvian, Milad},
  journal = {PRX Quantum},
  volume = {5},
  issue = {2},
  pages = {020355},
  numpages = {30},
  year = {2024},
  month = {Jun},
  publisher = {American Physical Society},
  doi = {10.1103/PRXQuantum.5.020355},
  url = {https://link.aps.org/doi/10.1103/PRXQuantum.5.020355}
}

@article{Mirrahimi_NJP_2014,
doi = {10.1088/1367-2630/16/4/045014},
url = {https://doi.org/10.1088/1367-2630/16/4/045014},
year = {2014},
month = {apr},
publisher = {IOP Publishing},
volume = {16},
number = {4},
pages = {045014},
author = {Mirrahimi, Mazyar and Leghtas, Zaki and Albert, Victor V and Touzard, Steven and Schoelkopf, Robert J and Jiang, Liang and Devoret, Michel H},
title = {Dynamically protected cat-qubits: a new paradigm for universal quantum computation},
journal = {New Journal of Physics},
}

@article{Morello-cats-2025,
	author = {Yu, Xi and Wilhelm, Benjamin and Holmes, Danielle and Vaartjes, Arjen and Schwienbacher, Daniel and Nurizzo, Martin and Kringh{\o}j, Anders and Blankenstein, Mark R. van and Jakob, Alexander M. and Gupta, Pragati and Hudson, Fay E. and Itoh, Kohei M. and Murray, Riley J. and Blume-Kohout, Robin and Ladd, Thaddeus D. and Anand, Namit and Dzurak, Andrew S. and Sanders, Barry C. and Jamieson, David N. and Morello, Andrea},
	doi = {10.1038/s41567-024-02745-0},
	journal = {Nature Physics},
	number = {3},
	pages = {362--367},
	title = {Schr{\"o}dinger cat states of a nuclear spin qudit in silicon},
	url = {https://doi.org/10.1038/s41567-024-02745-0},
	volume = {21},
	year = {2025},
}

@article{Morello-Nature-2022,
	author = {M{\k a}dzik, Mateusz T. and Asaad, Serwan and Youssry, Akram and Joecker, Benjamin and Rudinger, Kenneth M. and Nielsen, Erik and Young, Kevin C. and Proctor, Timothy J. and Baczewski, Andrew D. and Laucht, Arne and Schmitt, Vivien and Hudson, Fay E. and Itoh, Kohei M. and Jakob, Alexander M. and Johnson, Brett C. and Jamieson, David N. and Dzurak, Andrew S. and Ferrie, Christopher and Blume-Kohout, Robin and Morello, Andrea},
	doi = {10.1038/s41586-021-04292-7},
	journal = {Nature},
	number = {7893},
	pages = {348--353},
	title = {Precision tomography of a three-qubit donor quantum processor in silicon},
	url = {https://doi.org/10.1038/s41586-021-04292-7},
	volume = {601},
	year = {2022},
}

@article{Pfaff-NatPhys-2013,
	author = {Pfaff, Wolfgang and Taminiau, Tim H. and Robledo, Lucio and Bernien, Hannes and Markham, Matthew and Twitchen, Daniel J. and Hanson, Ronald},
	doi = {10.1038/nphys2444},
	journal = {Nature Physics},
	number = {1},
	pages = {29--33},
	title = {Demonstration of entanglement-by-measurement of solid-state qubits},
	url = {https://doi.org/10.1038/nphys2444},
	volume = {9},
	year = {2013},
}

@article{Pla-Nature-2013,
	author = {Pla, Jarryd J. and Tan, Kuan Y. and Dehollain, Juan P. and Lim, Wee H. and Morton, John J. L. and Zwanenburg, Floris A. and Jamieson, David N. and Dzurak, Andrew S. and Morello, Andrea},
	doi = {10.1038/nature12011},
	journal = {Nature},
	number = {7445},
	pages = {334--338},
	title = {High-fidelity readout and control of a nuclear spin qubit in silicon},
	url = {https://doi.org/10.1038/nature12011},
	volume = {496},
	year = {2013},
}

@article{Puri-2017,
	author = {Puri, Shruti and Boutin, Samuel and Blais, Alexandre},
	doi = {10.1038/s41534-017-0019-1},
	journal = {npj Quantum Information},
	number = {1},
	pages = {18},
	title = {Engineering the quantum states of light in a Kerr-nonlinear resonator by two-photon driving},
	url = {https://doi.org/10.1038/s41534-017-0019-1},
	volume = {3},
	year = {2017},
}

@article{puri2020biaspreserving,
  author  = {Shruti Puri and Lucas St-Jean and Jonathan A. Gross and Alexander Grimm and Nicolas E. Frattini and Pavithran S. Iyer and Anirudh Krishna and Steven Touzard and Liang Jiang and Alexandre Blais and Steven T. Flammia and S. M. Girvin},
  title   = {Bias-preserving gates with stabilized cat qubits},
  journal = {Science Advances},
  volume  = {6},
  number  = {34},
  pages   = {eaay5901},
  year    = {2020},
  doi     = {10.1126/sciadv.aay5901}
}

@article{Qing-PNAS-2026,
author = {Bingcheng Qing  and Ahmed Hajr  and Ke Wang  and Gerwin Koolstra  and Long B. Nguyen  and Jordan Hines  and Irwin Huang  and Bibek Bhandari  and Larry Chen  and Ziqi Kang  and Christian Jünger  and Noah Goss  and Nikitha Jain  and Hyunseong Kim  and Kan-Heng Lee  and Akel Hashim  and Nicholas E. Frattini  and Zahra Pedramrazi  and Justin Dressel  and Andrew N. Jordan  and David I. Santiago  and Irfan Siddiqi},
title = {Quantum benchmarking of high-fidelity noise-biased operations on a detuned Kerr-cat qubit},
journal = {Proceedings of the National Academy of Sciences},
volume = {123},
number = {5},
pages = {e2520479123},
year = {2026},
doi = {10.1073/pnas.2520479123},
URL = {https://www.pnas.org/doi/abs/10.1073/pnas.2520479123},
eprint = {https://www.pnas.org/doi/pdf/10.1073/pnas.2520479123}
}

@article{Raussendorf-Briegel-2001,
  title = {A One-Way Quantum Computer},
  author = {Raussendorf, Robert and Briegel, Hans J.},
  journal = {Phys. Rev. Lett.},
  volume = {86},
  issue = {22},
  pages = {5188--5191},
  numpages = {0},
  year = {2001},
  month = {May},
  publisher = {American Physical Society},
  doi = {10.1103/PhysRevLett.86.5188},
  url = {https://link.aps.org/doi/10.1103/PhysRevLett.86.5188}
}

@article{reglade2024catcontrol,
  author  = {Ulysse R{\'e}glade and Adrien Bocquet and Ronan Gautier and Joachim Cohen and Antoine Marquet and Emanuele Albertinale and Natalia Pankratova and Mattis Hall{\'e}n and Felix Rautschke and Lev-Arcady Sellem and Pierre Rouchon and Alain Sarlette and Mazyar Mirrahimi and Philippe Campagne-Ibarcq and Rapha{\"e}l Lescanne and S{\'e}bastien Jezouin and Zaki Leghtas},
  title   = {Quantum control of a cat qubit with bit-flip times exceeding ten seconds},
  journal = {Nature},
  volume  = {629},
  pages   = {778--783},
  year    = {2024},
  doi     = {10.1038/s41586-024-07294-3}
}

@misc{rousseau2025,
      title={Enhancing dissipative cat qubit protection by squeezing}, 
      author={Rémi Rousseau and Diego Ruiz and Emanuele Albertinale and Pol d'Avezac and Danielius Banys and Ugo Blandin and Nicolas Bourdaud and Giulio Campanaro and Gil Cardoso and Nathanael Cottet and Charlotte Cullip and Samuel Deléglise and Louise Devanz and Adam Devulder and Antoine Essig and Pierre Février and Adrien Gicquel and Élie Gouzien and Antoine Gras and Jérémie Guillaud and Efe Gümüş and Mattis Hallén and Anissa Jacob and Paul Magnard and Antoine Marquet and Salim Miklass and Théau Peronnin and Stéphane Polis and Felix Rautschke and Ulysse Réglade and Julien Roul and Jeremy Stevens and Jeanne Solard and Alexandre Thomas and Jean-Loup Ville and Pierre Wan-Fat and Raphaël Lescanne and Zaki Leghtas and Joachim Cohen and Sébastien Jezouin and Anil Murani},
      year={2025},
      eprint={2502.07892},
      archivePrefix={arXiv},
      primaryClass={quant-ph},
      url={https://arxiv.org/abs/2502.07892},}

@article{Ruiz-LDPC-2025,
	author = {Ruiz, Diego and Guillaud, J{\'e}r{\'e}mie and Leverrier, Anthony and Mirrahimi, Mazyar and Vuillot, Christophe},
	doi = {10.1038/s41467-025-56298-8},
	journal = {Nature Communications},
	number = {1},
	pages = {1040},
	title = {LDPC-cat codes for low-overhead quantum computing in 2D},
	url = {https://doi.org/10.1038/s41467-025-56298-8},
	volume = {16},
	year = {2025},
}

@article{Ruiz-npj-2026,
	author = {Ruiz, Diego and Guillaud, J{\'e}r{\'e}mie and Vuillot, Christophe and Mirrahimi, Mazyar},
	doi = {10.1038/s41534-026-01197-z},
	journal = {npj Quantum Information},
	number = {1},
	pages = {53},
	title = {Unfolded distillation: very low-cost magic state preparation for biased-noise qubits},
	url = {https://doi.org/10.1038/s41534-026-01197-z},
	volume = {12},
	year = {2026},
}

@misc{Ruiz2026_Review,
    author = {Ruiz, Diego and Guillaud, Jérémie and Vuillot, Christophe and Mirrahimi, Mazyar},
    title = {Faut-tolerant quantum computing with biased-noise qubits},
    year = {2026},
    note = {In preparation}
    }

@misc{shanahan2026,
      title={Elevator Codes: Concatenation for resource-efficient quantum memory under biased noise}, 
      author={Peter Shanahan and Diego Ruiz},
      year={2026},
      eprint={2601.10786},
      archivePrefix={arXiv},
      primaryClass={quant-ph},
      url={https://arxiv.org/abs/2601.10786}, 
}

@inproceedings{shor1996ftqc,
  author    = {Peter W. Shor},
  title     = {Fault-Tolerant Quantum Computation},
  booktitle = {Proceedings of the 37th Annual Symposium on Foundations of Computer Science (FOCS)},
  pages     = {56--65},
  year      = {1996},
  publisher = {IEEE Computer Society},
  address   = {Washington, DC}
}

@article{Taminiau-NatComm-2016,
	author = {Cramer, J. and Kalb, N. and Rol, M. A. and Hensen, B. and Blok, M. S. and Markham, M. and Twitchen, D. J. and Hanson, R. and Taminiau, T. H.},
	doi = {10.1038/ncomms11526},
	journal = {Nature Communications},
	number = {1},
	pages = {11526},
	title = {Repeated quantum error correction on a continuously encoded qubit by real-time feedback},
	url = {https://doi.org/10.1038/ncomms11526},
	volume = {7},
	year = {2016},
}

@article{Taminiau-Nature-2022,
	author = {Abobeih, M. H. and Wang, Y. and Randall, J. and Loenen, S. J. H. and Bradley, C. E. and Markham, M. and Twitchen, D. J. and Terhal, B. M. and Taminiau, T. H.},
	doi = {10.1038/s41586-022-04819-6},
	journal = {Nature},
	number = {7916},
	pages = {884--889},
	title = {Fault-tolerant operation of a logical qubit in a diamond quantum processor},
	url = {https://doi.org/10.1038/s41586-022-04819-6},
	volume = {606},
	year = {2022},
}

@article{Taminiau-PRL-2012,
  title = {Detection and Control of Individual Nuclear Spins Using a Weakly Coupled Electron Spin},
  author = {Taminiau, T. H. and Wagenaar, J. J. T. and van der Sar, T. and Jelezko, F. and Dobrovitski, V. V. and Hanson, R.},
  journal = {Phys. Rev. Lett.},
  volume = {109},
  issue = {13},
  pages = {137602},
  numpages = {5},
  year = {2012},
  month = {Sep},
  publisher = {American Physical Society},
  doi = {10.1103/PhysRevLett.109.137602},
  url = {https://link.aps.org/doi/10.1103/PhysRevLett.109.137602}
}

@article{OSullivan-NatPhys-2025,
	author = {O'Sullivan, James and Travesedo, Jaime and Pallegoix, Louis and Huang, Zhiyuan W. and Hogan, Patrick and May, Alexandre S. and Yavkin, Boris and Lin, Sen and Liu, Ren-Bao and Chaneliere, Thierry and Bertaina, Sylvain and Goldner, Philippe and Est{\`e}ve, Daniel and Vion, Denis and Abgrall, Patrick and Bertet, Patrice and Flurin, Emmanuel},
	date = {2025/11/01},
	doi = {10.1038/s41567-025-03049-7},
	id = {O'Sullivan2025},
	isbn = {1745-2481},
	journal = {Nature Physics},
	number = {11},
	pages = {1794--1800},
	title = {Individual solid-state nuclear spin qubits with coherence exceeding seconds},
	url = {https://doi.org/10.1038/s41567-025-03049-7},
	volume = {21},
	year = {2025}}

@article{Travesedo-2025,
author = {Jaime Travesedo  and James O'Sullivan  and Louis Pallegoix  and Zhiyuan W. Huang  and Patrick Hogan  and Philippe Goldner  and Thierry Chaneliere  and Sylvain Bertaina  and Daniel Estève  and Patrick Abgrall  and Denis Vion  and Emmanuel Flurin  and Patrice Bertet },
title = {All-microwave spectroscopy and polarization of individual nuclear spins in a solid},
journal = {Science Advances},
volume = {11},
number = {10},
pages = {eadu0581},
year = {2025},
doi = {10.1126/sciadv.adu0581},
URL = {https://www.science.org/doi/abs/10.1126/sciadv.adu0581},
eprint = {https://www.science.org/doi/pdf/10.1126/sciadv.adu0581}
}

@article{tuckett2018ultrahigh,
  author  = {David K. Tuckett and Stephen D. Bartlett and Steven T. Flammia},
  title   = {Ultrahigh Error Threshold for Surface Codes with Biased Noise},
  journal = {Phys. Rev. Lett.},
  volume  = {120},
  pages   = {050505},
  year    = {2018},
  doi     = {10.1103/PhysRevLett.120.050505}
}

@article{Vandersypen-Science-2018,
author = {N. Samkharadze  and G. Zheng  and N. Kalhor  and D. Brousse  and A. Sammak  and U. C. Mendes  and A. Blais  and G. Scappucci  and L. M. K. Vandersypen },
title = {Strong spin-photon coupling in silicon},
journal = {Science},
volume = {359},
number = {6380},
pages = {1123-1127},
year = {2018},
doi = {10.1126/science.aar4054},
URL = {https://www.science.org/doi/abs/10.1126/science.aar4054},
eprint = {https://www.science.org/doi/pdf/10.1126/science.aar4054}
}

@article{Wolfowicz_2016,
doi = {10.1088/1367-2630/18/2/023021},
url = {https://doi.org/10.1088/1367-2630/18/2/023021},
year = {2016},
month = {feb},
publisher = {IOP Publishing},
volume = {18},
number = {2},
pages = {023021},
author = {Wolfowicz, Gary and Mortemousque, Pierre-André and Guichard, Roland and Simmons, Stephanie and Thewalt, Mike L W and Itoh, Kohei M and Morton, John J L},
title = {29Si nuclear spins as a resource for donor spin qubits in silicon},
journal = {New Journal of Physics}
}

\appendix

\section{CZ-based approaches}
\label{sec:CZ}

In this appendix, we compare the performance of two phase-flip repetition code schemes based on bias-preserving C$Z$ gates: the first one employs such gates to perform QND multi-Z measurements at the logical level of a repetition code~\cite{aliferis2008biased,brooks-preskill-2013}, and the second one uses the C\(Z\) gates to perform QND multi-Z measurements directly at the physical level of a biased-noise qubit.

\begin{figure}[ht]
    \centering
    \includegraphics[width=.9\columnwidth]{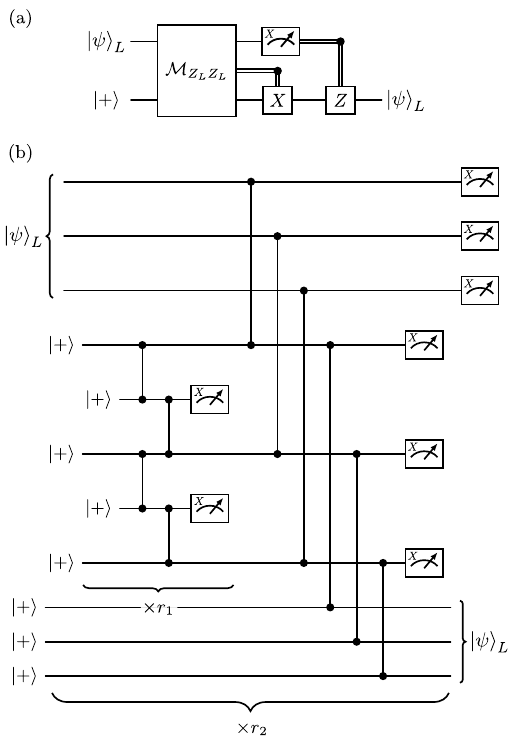}
    \caption{(a) Knill-type quantum error correction using a joint $ZZ$ logical measurement. The circuit teleports the encoded logical qubit $\ket{\psi}_L$ from the first logical block to the second. The transversal $X$ measurement on the first qubit allows the reconstruction of $X$ stabilizers and thus the detection of phase-flip errors. (b) Physical implementation of the circuit in (a). The measurement of the high-weight $Z$ operator is performed using bias-preserving $\text C Z$ gates. The first step is to prepare a $\ket{GHZ}$ state. This is performed by preparing $n$ data qubits in $\ket{+}^{\otimes n}$ and measuring neighbouring $ZZ$ stabilizers of the state using $n-1$ ancilla qubits (here $n=3$). This measurement is noisy and is thus repeated $r_1$ times. This $\ket{GHZ}$ state is then used to measure the weight-$2n$ $Z$ operator with the application of transversal $\text C Z$ gates. As this step is noisy as well, it is repeated $r_2$ times, leading to a total of $r_1 r_2$ rounds for one teleportation.}
    \label{fig:Aliferis-Knill}
\end{figure}

More precisely, the first scheme, proposed in~\cite{brooks-preskill-2013}, is adapted from Knill-type quantum error correction and consists in teleporting the quantum information between two blocks, allowing for syndrome measurements, as shown in Figure~\ref{fig:Aliferis-Knill}. The teleportation involves a logical $Z_L Z_L$ measurement, performed with a $\ket{GHZ}$ state. As this measurement is noisy, it is repeated $r_2$ times, while the $\ket{GHZ}$ preparation itself requires repeated physical $ZZ$ measurement. Thus, $r_1 \times r_2$ rounds are required in total to extract the $X$-stabilizers of the repetition code. However, during this lengthy measurement, phase-flip errors accumulate on the data qubits of the repetition code, such that overall the scheme has no threshold even in the presence of phase-flip errors only.

\begin{figure}[ht]
    \centering
    \includegraphics[width=.9\columnwidth]{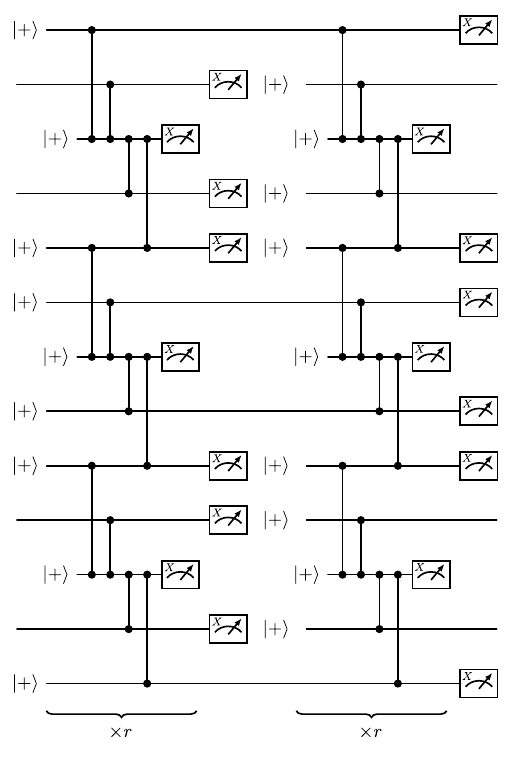}
    \caption{Phase-flip repetition code ($d=4$ here) where $\mathcal M_{XX}$ operations are compiled according to Figure~\ref{fig:CX}. The weight-4 $Z$ measurements are repeated $r$ times for fault-tolerance.}
    \label{fig:Z_physical_rep_code}
\end{figure}
The second scheme uses the bias-preserving C\(Z\) gates to perform physical \(M_{Z^{\otimes 4}}\) measurements, which in turn let us measure the stabilizers of the repetition code, as shown in Figure~\ref{fig:Z_physical_rep_code}. In contrast with the first scheme, the weight of the $Z$ measurements stays constant and does not grow with the distance $d$. Overall, this method is more efficient than the first scheme as less rounds are required to achieve high-fidelity \(M_{Z^{\otimes 4}}\) measurements.

\begin{figure}[ht]
    \centering
    \includegraphics[width=\columnwidth]{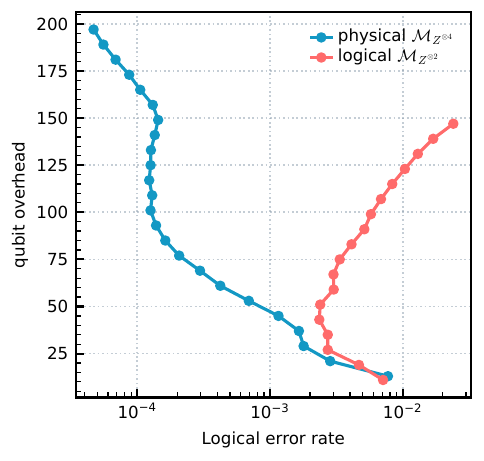}
    \caption{Qubit overhead as a function of the logical error rate of the two schemes presented in Appendix~\ref{sec:CZ}. The logical error rate is computed per round of $X$-stabilizer measurement and for a physical phase-flip error rate per operation of $p_z=10^{-3}$.}
    \label{fig:CZ-numerics}
\end{figure}

In Figure~\ref{fig:CZ-numerics}, we compare the qubit overhead required to reach a given logical error rate at a physical error rate per qubit and per operation of $p_z = 10^{-3}$. The logical error rate is reported per round of $X$-stabilizer measurements, and the optimal $r_1$, $r_2$ or $r$ that minimizes the logical error rate for each qubit count is selected. Single-qubit gates and idle locations are followed by a phase-flip error occurring with probability $p_z$, while two-qubit C$Z$ gates are subject to $IZ$, $ZI$, and $ZZ$ errors, each occurring with probability \(\frac{2p_z}{3}\). As expected, whereas the first scheme quickly saturates, the second scheme enables us to explore a much lower logical error rate regime.

\section{Numerical simulations}
\label{sec:sc_simulations}
\begin{table}[tp!]
\centering
\caption{Comparison of the circuit-level noise channels for the two
simulated codes.  Each row corresponds to one operation type.
We define $p_\mathrm{cx} = 2p_z(1+\eta^{-1})$. \(\texttt{DEPOLARIZE\_2}(p)\) designate the two-qubit depolarizing channel of parameter \(p\), \(\texttt{PAULI\_1}(p_x,p_y,p_z)\) the single qubit Pauli channel of parameters \(p_x\), \(p_y\) and \(p_z\) corresponding to \(X\), \(Y\) and \(Z\) error rates, \(\texttt{X\_ERROR}(p)\) \(X\) error with probability \(p\) and \(\texttt{Z\_ERROR}(p)\) \(Z\) error with probability \(p\).}
\label{tab:noise_models}
\begin{tabular}{|l|l|l|}
\hline
\textbf{Operation} & \textbf{Rotated surface code} & \textbf{XZZX + MZ$^4$} \\
\hline
CX/CZ & $\texttt{DEPOLARIZE\_2}(p_\mathrm{cx})$ & $\texttt{PAULI\_1}(\frac{p_z}{2\eta},\frac{p_z}{2\eta},p_z)$ \\\hline
Idle & \multicolumn{2}{c|}{$\texttt{PAULI\_1}(\frac{p_z}{2\eta},\frac{p_z}{2\eta},p_z)$} \\\hline
MZ & $\texttt{X\_ERROR}(\frac{p_z}{\eta})$ & MZ$^4$ assignment err.\ $\frac{p_z}{\eta}$ \\
MX & $\texttt{Z\_ERROR}(p_z)$ & MX assignment err.\ $p_z$ \\\hline
Reset & \multicolumn{2}{c|}{$\texttt{PAULI\_1}(\frac{p_z}{2\eta},\frac{p_z}{2\eta},p_z)$} \\
\hline
\end{tabular}
\end{table}

In this appendix, we describe the circuit-level noise models and simulation
parameters used for the overhead comparison of Section~\ref{sec:performances}.
All circuits are constructed with \texttt{stim}~\cite{gidney2021stim} and
decoded with \texttt{PyMatching}~\cite{higgott2022pymatching}.

\subsection{Circuit-level noise models}

Table~\ref{tab:noise_models} summarises the noise channels applied to each
operation in the two simulation setups.  Both models are parameterised by
$p_z$ and $\eta$; the rotated surface code uses a depolarizing CX with rate
$p_\mathrm{cx} = 2p_z(1+\eta^{-1})$ while all single-qubit channels retain the
native noise bias.

The rotated surface code noise model uses a depolarizing channel
$\texttt{DEPOLARIZE\_2}(p_\mathrm{cx})$ after every CX gate, while
all single-qubit operations (resets and idle time steps)
apply the bias-preserving Pauli channel
$\texttt{PAULI\_1}(\frac{p_z}{2\eta},\frac{p_z}{2\eta},p_z)$.
Z-basis measurement outcomes are flipped with probability $\frac{p_z}{\eta}$
and X-basis measurements with probability $p_z$.
This model captures the scenario where single-qubit operations
inherit the native noise bias of the hardware but the two-qubit
gate is non-bias-preserving.

The XZZX$+$MZ$^4$ noise model applies an asymmetric Pauli channel
$\texttt{PAULI\_1}(\frac{p_z}{2\eta},\;\frac{p_z}{2\eta},\;p_z)$ after every layer
(gates, measurements, resets, and idle time steps) to all alive qubits.
The MZ$^4$ measurements (implemented as \texttt{MPP} $Z^{\otimes 4}$ in stim)
suffer an assignment error rate $\frac{p_z}{\eta}$, dominated by
bit-flip processes, while the MX teleportation measurements flip with
probability $p_z$.  These rates reflect the expected noise hierarchy of a
bias-preserving architecture: $Z$-type errors occur at rate $p_z$ while
$X/Y$-type errors are suppressed by $\eta$.

\subsection{Simulation parameters}

\textit{Rotated surface code.}
We simulate the rotated $d\times d$ surface code ($n = 2d^2-1$ qubits) for
$d \in \{3,5,7,9,11,13\}$ and $\mathrm{rounds} = d$, with both $Z$- and
$X$-memory experiments.  The bias is set to $\eta = 10^6$ (effectively
infinite, so that $p_\mathrm{cx} \approx 2p_z$) and
$p_z$ is swept over 14 values from $5\times 10^{-4}$ to $3\times 10^{-2}$.
Statistics are collected up to $5\times 10^6$~shots or 500~logical errors per task,
whichever comes first.
\begin{figure}[h!]
    \centering
    \includegraphics[width=0.75\columnwidth]{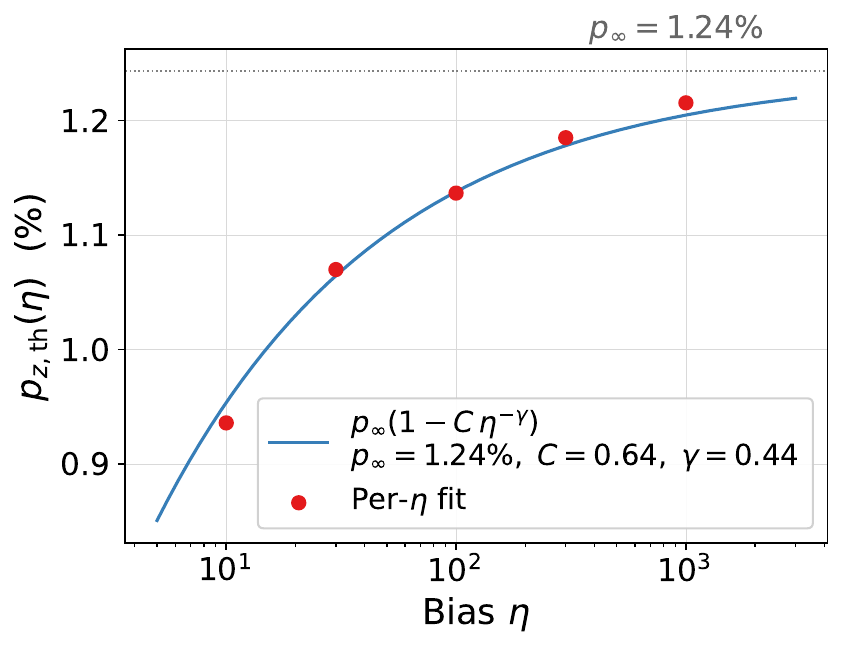}
    \caption{%
        Effective threshold $p_{z,\mathrm{th}}(\eta)$ for the $\ket{+}$ memory.
        Red dots: per-$\eta$ fitted values.
        Blue curve: global parametrization
        $p_{z,\mathrm{th}}(\eta) = p_\infty(1 - C\,\eta^{-\gamma})$
        [Eq.~\eqref{eq:pth_eta}].
        The dotted line marks the asymptotic value $p_\infty \approx 1.25\%$.
    }
    \label{fig:pth_vs_eta}
\end{figure}

\textit{XZZX surface code.}
We simulate the non-rotated XZZX surface code with the alternating MZ$^4$ schedule
for $d_x \in \{3,5\}$, $d_z \in \{7,9,11,13,15,17\}$, and
$\eta \in \{10,30,100,300,1000\}$, with both $\ket{+}$ and $\ket{0}$ memory
experiments.  Three physical error rates are used: $p_z \in \{10^{-3}, 3\times 10^{-3}, 5\times 10^{-3}\}$.
The number of rounds per shot is $\max(d_x, d_z)$.
Per-cycle logical error rates are obtained as
$p_L = 1 - (1 - p_L^\mathrm{shot})^{1/\mathrm{rounds}}$.
Statistics are collected up to $5\times 10^9$~shots or 300~logical errors per
task for the hardest parameter points; data points with fewer than 50~observed
errors are excluded from fits.

\subsection{Fitted sub-threshold ansätze}
\label{sec:fitting}

\textit{Rotated surface code.} The $X$- and $Z$-memory logical error rates are fitted separately to
\begin{equation}
    p_L^\mathrm{SC} = A_\mathrm{SC}\,
    \left(\frac{p_\mathrm{cx}}{p_\mathrm{th}}\right)^{(d+1)/2},
    \label{eq:ansatz_sc}
\end{equation}
where $p_\mathrm{cx} = 2p_z(1 + \eta^{-1})$.
The $X$-memory fit (limited by depolarized CX errors, and thus the
bottleneck at large $\eta$) yields
$A_\mathrm{SC}^{(X)} \approx 0.042$ and $p_\mathrm{th}^{(X)} \approx 0.46\%$.
The $Z$-memory fit yields
$A_\mathrm{SC}^{(Z)} \approx 0.016$ and $p_\mathrm{th}^{(Z)} \approx 0.68\%$.
The lower $X$-memory threshold compared to the standard literature value
$p_\mathrm{th} \approx 0.94\%$~\cite{fowler2012surface,dennis2002topological}
for circuit-level depolarising noise is expected: our model uses
$p_\mathrm{cx} = 2p_z(1+\eta^{-1})$---twice the standard depolarising rate---reflecting
the full error budget of a non-bias-preserving two-qubit gate.
The overhead comparison uses
$\mathrm{LER}_\mathrm{SC} = p_L^{(X)} + p_L^{(Z)}$, which is dominated by
the $X$-memory contribution at large $\eta$.

\textit{XZZX surface code.}
The $\ket{+}$ and $\ket{0}$ memory logical error rates are fitted to the
ansätze of Eqs.~\eqref{eq:ansatz_plus} and~\eqref{eq:ansatz_zero}.

\textit{$\ket{+}$ memory.}
Logical errors in the $\ket{+}$ memory experiment arise from $Z$-error chains spanning
the code in the $d_z$ direction.  The multiplicity of shortest-weight logical
representatives scales as $d_x$, motivating the ansatz
\begin{equation}
    p_L^{(+)} = A\, d_x \left(\frac{p_z}{p_{z,\mathrm{th}}(\eta)}\right)^{(d_z+1)/2},
    \tag{\ref{eq:ansatz_plus}}
\end{equation}
where the effective threshold $p_{z,\mathrm{th}}(\eta)$ captures the reduction in
threshold at finite bias due to subleading $X/Y$ errors.
We parametrize this dependence as
\begin{equation}
    p_{z,\mathrm{th}}(\eta) = p_\infty\!\left(1 - C\,\eta^{-\gamma}\right),
    \label{eq:pth_eta}
\end{equation}
so that $p_{z,\mathrm{th}} \to p_\infty$ as $\eta \to \infty$.
A global fit in log-space to all $\ket{+}$ data points with at least one observed
logical error yields $A \approx 0.042$, $p_\infty = 1.25\!\times\!10^{-2}$,
$C = 0.623$, $\gamma = 0.429$.
As $\eta \to \infty$, the threshold saturates to $p_\infty \approx 1.25\%$.
Figure~\ref{fig:pth_vs_eta} compares this parametrization to the per-$\eta$
fitted thresholds, confirming the quality of the fit.

\textit{$\ket{0}$ memory.}
The $\ket{0}$ memory is limited by $X$-error chains of length $d_x$ with multiplicity
scaling as $d_z$:
\begin{equation}
    p_L^{(0)} = A'\, d_z
    \left(\frac{p_z}{\eta^\beta\, p_{x,\mathrm{th}}}\right)^{\alpha (d_x+1)/2}.
    \tag{\ref{eq:ansatz_zero}}
\end{equation}
Here the exponent $\beta$ absorbs the $\eta$-dependence of the effective bit-flip
threshold (theory predicts $\beta = 1$ for pure scaling).
A global fit yields $A' \approx 0.039$, $p_{x,\mathrm{th}} = 5.3\!\times\!10^{-3}$,
$\beta = 0.685$, $\alpha = 1.264$.
Theory predicts $\alpha = 1$; the larger fitted value may reflect the limited
range of $d_x$ values ($d_x \in \{3,5\}$) and the fact that many data points at
high bias lie below the statistical floor ($p_L < 10^{-8}$).

\subsection{XZZX surface code layout and scheduling}
\label{ssec:xzzx_layout}

\begin{figure}[htbp]
    \centering
    \includegraphics[width=0.65\columnwidth]{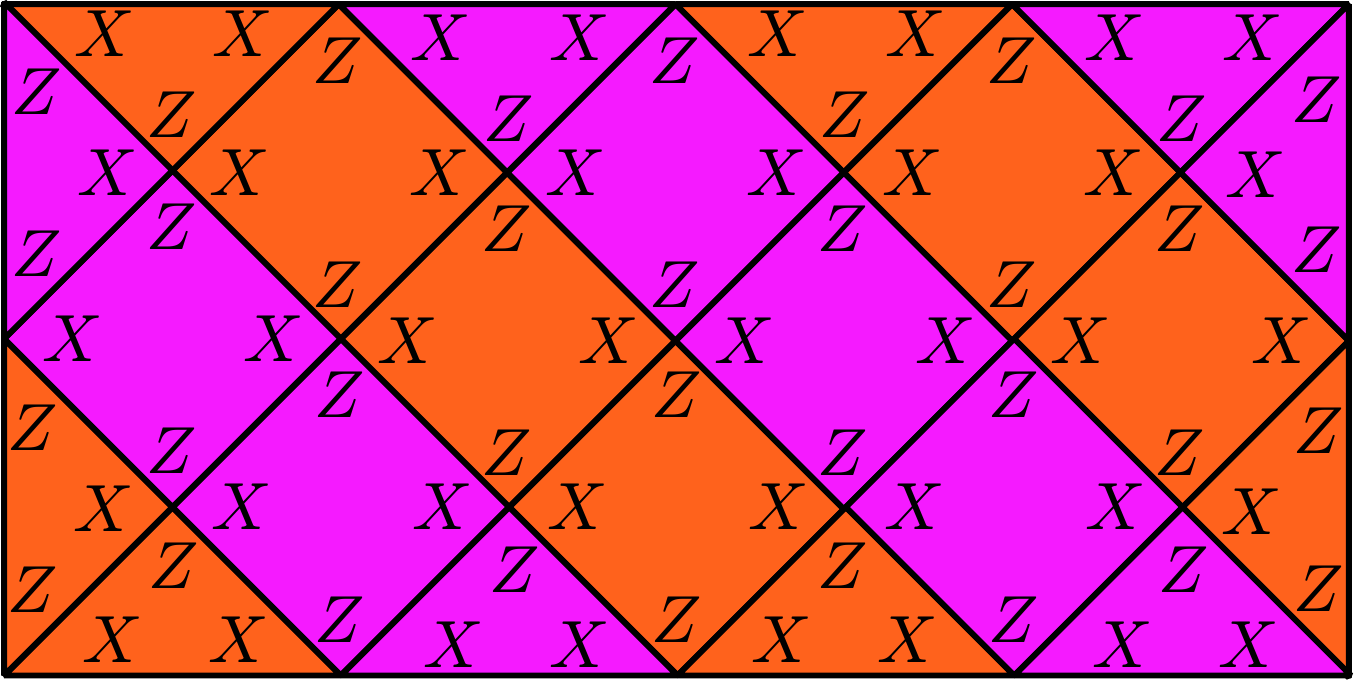}
    \caption{%
        Stabilizer partition for the XZZX surface code with alternating
        $M_{Z^4}$ measurements.  The two sets of stabilizers (orange and purple)
        are measured on alternate rounds.  Each vertex hosts two physical qubits.
    }
    \label{fig:XZZX_schedule_app}
\end{figure}

\begin{figure}[htbp]
    \centering
    \includegraphics[width=0.3\textwidth]{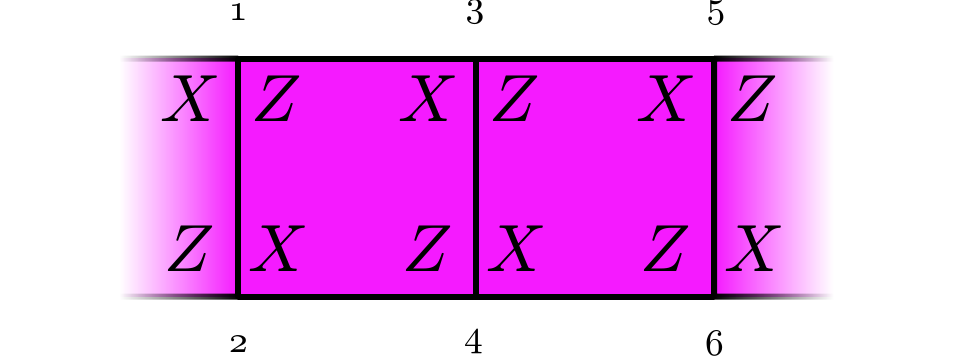}\\[1em]
    \raisebox{7em}{\resizebox{.6\linewidth}{!}{\begin{quantikz}[row sep=0.5em, column sep=0.5em]
        \lstick{$\vdots$}&\setwiretype{n}&\setwiretype{q}& \phase{}&&\cdots\setwiretype{n}&&\ghost{X}\setwiretype{q}& \\
        \lstick[2]{1} &\setwiretype{n}&\ctrl{2}\setwiretype{q}&&&\cdots\setwiretype{n}&\ghost{X}& \setwiretype{q}&\\
        &\setwiretype{n}\midstick{$\ket{+}$} &\setwiretype{q}&\ctrl{2}&&\cdots\setwiretype{n}&\ghost{X}&\setwiretype{q}&\\
        \lstick[2]{2} &\setwiretype{n}&\ctrl{-2}\setwiretype{q}&&& \gate[4, disable auto height][1][1]{\begin{matrix}
                M_{Z^4}\\ 	\downarrow\\s_1
        \end{matrix}}&\meterD{M_X}\rstick{\hspace{-.45em}$= m_1$} \\
        &\setwiretype{n}\midstick{$\ket{+}$} &\setwiretype{q}&\ctrl{-2}&&&&\gate{Z^{m_1}}&\\
        \lstick[2]{3} &\setwiretype{n}&\ctrl{2}\setwiretype{q}&&&& \meterD{M_X}\rstick{\hspace{-.45em}$= m_2$}\\
        &\setwiretype{n}\midstick{$\ket{+}$} &\setwiretype{q}&\ctrl{2}&&&&\gate{X^{s_1}Z^{m_2}}&\\
        \lstick[2]{4} &\setwiretype{n}&\ctrl{-2}\setwiretype{q}&&&\gate[4, disable auto height][1][1]{\begin{matrix}
                M_{Z^4}\\ 	\downarrow\\s_2
        \end{matrix}}& \meterD{M_X}\rstick{\hspace{-.45em}$= m_3$}\\
        &\setwiretype{n}\midstick{$\ket{+}$} &\setwiretype{q}&\ctrl{-2}&&&& \gate{Z^{m_3+s_1}}&\\
        \lstick[2]{5} &\setwiretype{n}&\ctrl{2}\setwiretype{q}&&&&\meterD{M_X}\rstick{\hspace{-.45em}$= m_4$}\\
        &\setwiretype{n}\midstick{$\ket{+}$} &\setwiretype{q}&\ctrl{2}&& &&\gate{X^{s_2}Z^{m_4}}& \\
        \lstick[2]{6} &\setwiretype{n}&\ctrl{-2}\setwiretype{q}&&&\cdots\setwiretype{n}&\ghost{X}& \setwiretype{q}&\\
        &\setwiretype{n}\midstick{$\ket{+}$} &\setwiretype{q}&\ctrl{-2}&&\cdots\setwiretype{n}&&\gate{Z^{s_2}} \setwiretype{q}&\\
        \lstick{$\vdots$}&\setwiretype{n}&\phase{}\setwiretype{q}&&&\cdots\setwiretype{n}&&\ghost{X}\setwiretype{q}& \\
    \end{quantikz}}}
    
    \caption{%
        Syndrome extraction circuit for one partition of the XZZX surface code
        using $M_{Z^4}$ measurements.
    }
    \label{fig:XZZX_MZZZZ_app}
\end{figure}

The XZZX surface code with alternating $M_{Z^4}$ syndrome extraction requires
partitioning the stabilizers into two sets that are measured on alternate rounds.
Figure~\ref{fig:XZZX_schedule_app} shows the partition: the two stabilizer sets
are colored in orange and purple.  Each vertex hosts two qubits (data and ancilla),
and the circuit of Figure~\ref{fig:XZZX_MZZZZ_app} is applied to each partition in
alternation.
At the boundaries there can be two types of three-body stabilizers, XZX or XZZ.
For the former the stabilizer measurement is straightforwardly adapted by only performing C\(Z\) towards the extra Z.
For the latter it is enough to introduce and extra qubit in state \(\ket{+}\) and measuring XZZX including this extra qubit.

\section{Alternated vs.\ simultaneous \texorpdfstring{$M_{Z^4}$}{MZ4} schedules}
\label{sec:alt_vs_sim}

In the main text, we use the \emph{alternated} schedule for syndrome extraction with $M_{Z^4}$ gates: data qubits are arranged on $2d$ sites and each round applies half of the $M_{Z^4}$ measurements, so that a full syndrome is obtained after two rounds (four time steps).
An alternative is the \emph{simultaneous} schedule, which applies all $M_{Z^4}$ measurements in a single round at the cost of extra ancilla qubits: $3d-2$ qubits total. One round then consists of only two time steps (the $M_{Z^4}$ step followed by $X$-measurements and data relocation).

\begin{figure}[ht]
    \centering
    \raisebox{9em}{\resizebox{.7\linewidth}{!}{\begin{quantikz}[row sep=0.5em]
			\lstick{$\ket{+}$} & \gate[4, disable auto height][1][1]{\begin{matrix}
					M_{Z^4}\\ 	\downarrow\\s_1
			\end{matrix}}&&\gate{X^{s_1}Z^{m_1}}& \\
			\lstick{$D_0$} &&& \meterD{M_X}\rstick{\hspace{-.45em}$= m_1$}\\
			\lstick{$D_1$} &&& \meterD{M_X}\rstick{\hspace{-.45em}$= m_2$}\\
			\lstick{$\ket{+}$} &&\gate[4, disable auto height][1][1]{\begin{matrix}
					M_{Z^4}\\ 	\downarrow\\s_2
			\end{matrix}}& \meterD{M_X}\rstick{\hspace{-.45em}$= m_3$}\\
			\lstick{$\ket{+}$} &&& \gate{X^{s_2}Z^{m_2+m_3}}&\\
			\lstick{$\ket{+}$} &&&\gate{X^{s_3}Z^{m_4+m_5}}& \\
			\lstick{$\ket{+}$} & \gate[4, disable auto height][1][1]{\begin{matrix}
					M_{Z^4}\\ 	\downarrow\\s_3
			\end{matrix}}&& \meterD{M_X}\rstick{\hspace{-.45em}$= m_4$}\\
			\lstick{$D_2$} &&& \meterD{M_X}\rstick{\hspace{-.45em}$= m_5$}\\
			\lstick{$D_3$} &&& \meterD{M_X}\rstick{\hspace{-.45em}$= m_6$}\\
			\lstick{$\ket{+}$} &&\gate[4, disable auto height][1][1]{\begin{matrix}
					M_{Z^4}\\ 	\downarrow\\s_4
			\end{matrix}}& \meterD{M_X}\rstick{\hspace{-.45em}$= m_7$}\\
			\lstick{$\ket{+}$} &&&\gate{X^{s_4}Z^{m_6+m_7}}& \\
			\lstick{$\ket{+}$} &&&\gate{Z^{m_8}}& \\
			\lstick{$D_4$} &&& \meterD{M_X}\rstick{\hspace{-.45em}$= m_8$}\\
		\end{quantikz}}}
        \\
    \includegraphics[width=0.7\linewidth]{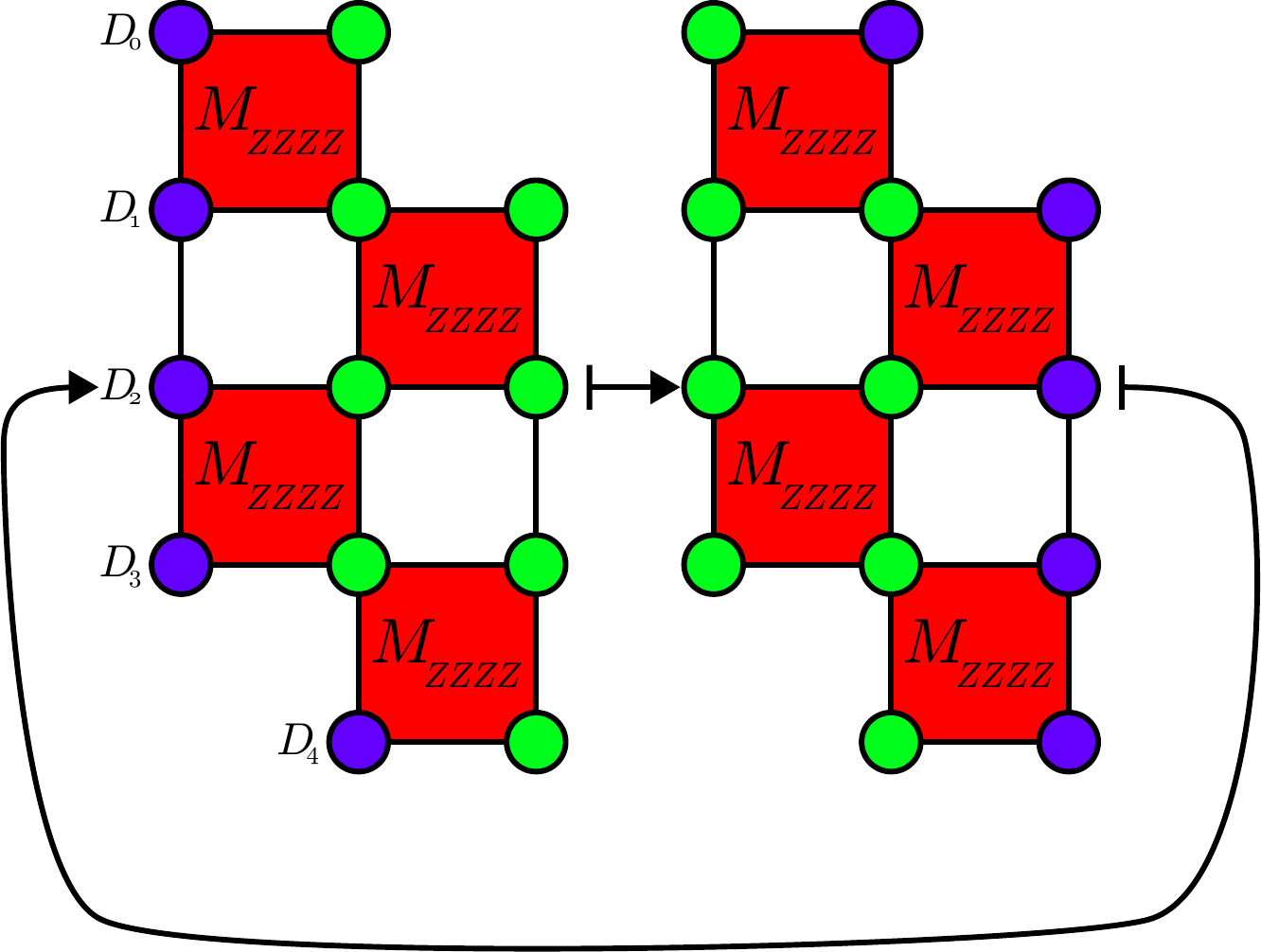}
        
    \caption{Circuit for the simultaneous $M_{Z^4}$ schedule with $d=5$.
    All $d-1$ stabilizers are measured in a single step using $3d-2$ qubits.
    The full syndrome is extracted in a single round (two time steps), compared to two rounds (four time steps) for the alternated schedule.}
    \label{fig:alt_vs_sim_circuit}
\end{figure}

Figure~\ref{fig:alt_vs_sim_circuit} shows the circuit for the simultaneous schedule.
Since all stabilizers are measured in every round, the effective noise per syndrome cycle is $p_{\rm eff}^{\rm sim} = 2\,p_z$ compared to $p_{\rm eff}^{\rm alt} = 4\,p_z$ for the alternated schedule.

\begin{figure*}[t]
    \centering
    \includegraphics[width=.9\textwidth]{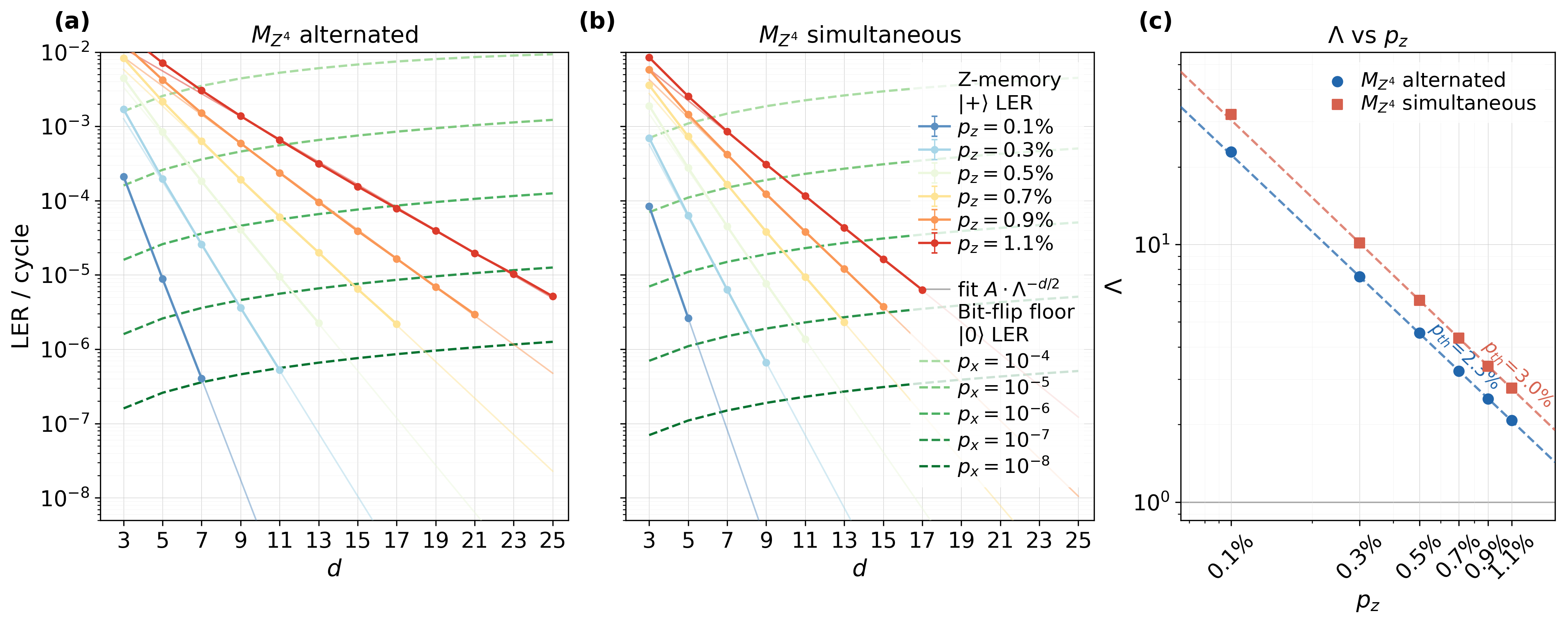}
    \caption{Comparison of the alternated and simultaneous $M_{Z^4}$ schedules.
    (a)~Logical error rate per cycle for the alternated schedule ($2d$ qubits).
    (b)~Same for the simultaneous schedule ($3d{-}2$ qubits).
    Solid lines: exponential fits $A\cdot\Lambda^{-d/2}$; dashed lines: bit-flip floor from the DEM.
    (c)~Suppression factor $\Lambda$ vs.\ $p_z$; the simultaneous schedule has a higher threshold ($p_{\rm th} \approx 3.0\%$) than the alternated one ($p_{\rm th} \approx 2.3\%$).}
    \label{fig:alt_vs_sim_combined}
\end{figure*}

\textit{Phase-flip threshold.}
Figure~\ref{fig:alt_vs_sim_combined} shows the logical error rate per cycle as a function of code distance for both schedules [panels~(a) and~(b)], together with the extracted suppression factor $\Lambda$ [panel~(c)].
Because the effective noise per round is halved, the simultaneous schedule exhibits a higher threshold: $p_{\rm th}^{\rm sim} \approx 3.0\%$ vs.\ $p_{\rm th}^{\rm alt} \approx 2.3\%$.
Both fits have slope $\approx -1$, consistent with $\Lambda \propto p_z / p_{\rm th}$.
Note that the threshold ratio $p_{\rm th}^{\rm sim}/p_{\rm th}^{\rm alt} \approx 1.3$ is less than the na\"ive factor of~$2$ expected from halving the number of time steps per round.
This is because the simultaneous schedule introduces additional ancilla qubits that are also subject to noise: in the teleportation step, the data qubit is now teleported through two $X$-measurements (from two adjacent $M_{Z^4}$ operations), each of which can imprint a $Z$ error on the outgoing qubit.
In the alternated schedule, each data qubit participates in a single $M_{Z^4}$ per round, so only one measurement outcome feeds into the teleportation correction.

\begin{figure}[t]
    \centering
    \includegraphics[width=0.85\columnwidth]{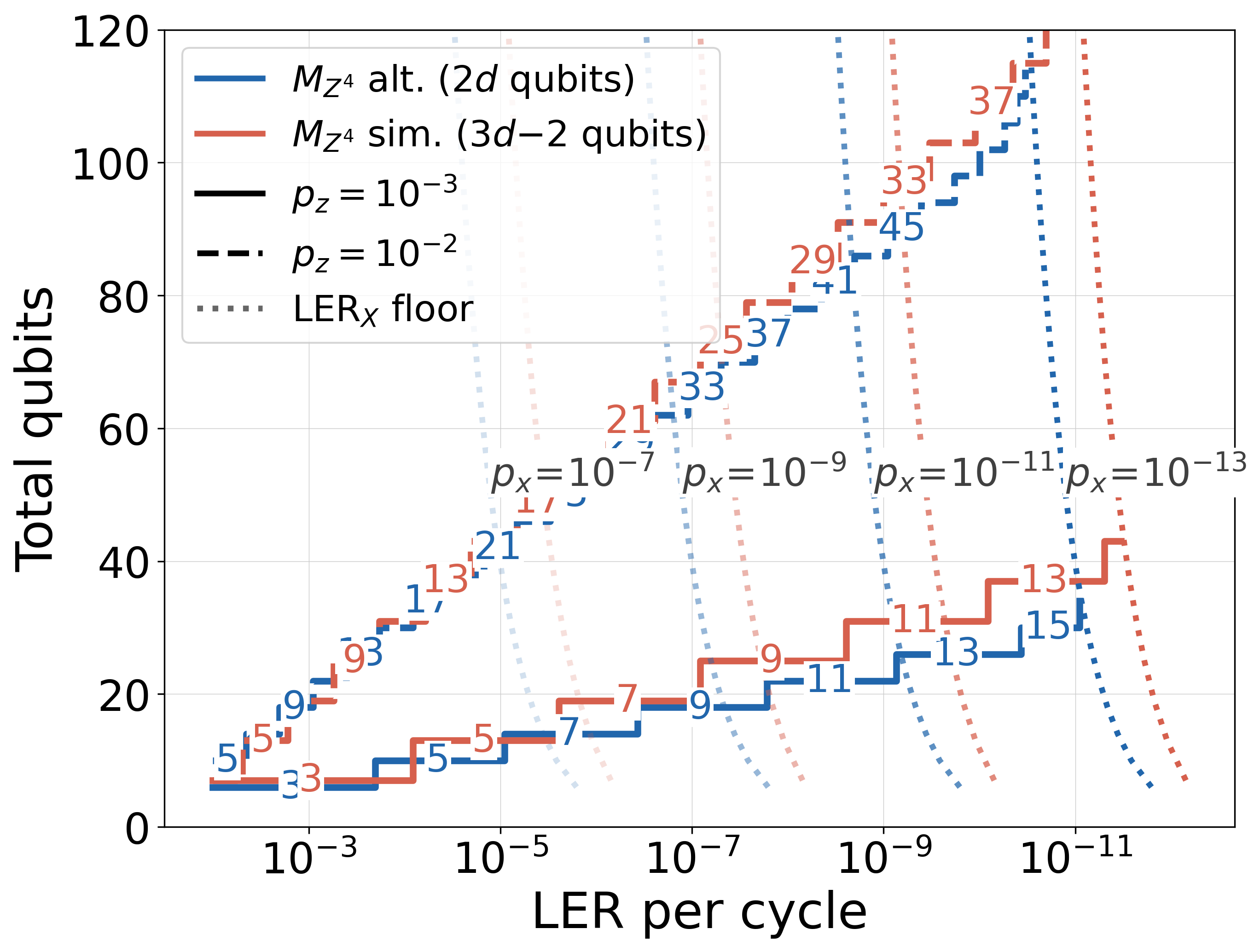}
    \caption{Total qubit overhead vs.\ achievable logical error rate per cycle for the alternated ($2d$ qubits, blue) and simultaneous ($3d{-}2$ qubits, red) schedules.
    Solid lines: $p_z = 10^{-3}$; dashed: $p_z = 10^{-2}$.
    Opacity encodes $p_x$ (faint: $10^{-7}$, opaque: $10^{-13}$).
    Dotted diagonal lines show the bit-flip (LER$_X$) floor.
    At moderate target LER, the two schedules have comparable overhead; the simultaneous schedule gains at very low target LER owing to its higher $\Lambda$.}
    \label{fig:alt_vs_sim_overhead}
\end{figure}

\begin{figure}[t]
    \centering
    \vspace{2em}
    \raisebox{4em}{\resizebox{.45\linewidth}{!}{\begin{quantikz}[row sep=0.5em]
			\lstick{$\ket{+}$} & \gate[4, disable auto height][1][1]{\begin{matrix}
					M_{Z^4}\\ 	\downarrow\\s_1
			\end{matrix}}&&\gate{Z^{m_1}}&\\
			\lstick{$D_0$} &&&\meterD{M_X}&\setwiretype{n} \midstick{\hspace{-2.3em}$=m_1$}\\
			\lstick{$\ket{+}$} &&\gate[4, disable auto height][1][1]{\begin{matrix}
					M_{Z^4}\\ 	\downarrow\\s_2
			\end{matrix}}&\gate{Z^{m_2}}&\\
			\lstick{$D_1$} &&&\meterD{M_X} &\setwiretype{n}\midstick{\hspace{-2.3em}$=m_2$}\\
			\lstick{$\ket{+}$} & \gate[4, disable auto height][1][1]{\begin{matrix}
					M_{Z^4}\\ 	\downarrow\\s_3
			\end{matrix}}&&\gate{Z^{m_3}}&\\
			\lstick{$D_2$} &&&\meterD{M_X} &\setwiretype{n}\midstick{\hspace{-2.3em}$=m_3$}\\
			\lstick{$\ket{+}$} &&&\gate{Z^{m_4}}&\\
			\lstick{$D_3$} &&&\meterD{M_X}&\setwiretype{n}\midstick{\hspace{-2.3em}$=m_4$}\\
		\end{quantikz}}}
        \hfil
        \raisebox{0em}{\includegraphics[width=0.35\linewidth]{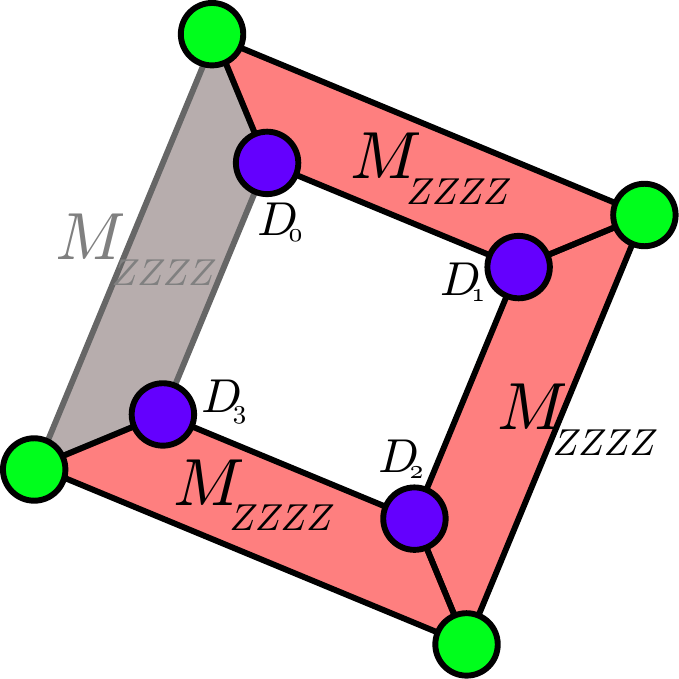}}
    \caption{Implementation of an \(M_{X^{\otimes 4}}\) \(M_{Z^{\otimes 4}}\) readouts, see main text.}
    \label{fig:MX4}
\end{figure}

\textit{Qubit overhead comparison.}
Despite the higher per-round $\Lambda$, the simultaneous schedule requires $3d-2$ qubits compared to $2d$ for the alternated schedule. This partially cancels the threshold advantage when comparing the total qubit count required to reach a given target logical error rate, see Figure~\ref{fig:alt_vs_sim_overhead}.
The figure shows the overhead for two values of $p_z$ (solid: $p_z = 10^{-3}$, dashed: $p_z = 10^{-2}$), with transparency encoding the bit-flip rate $p_x$ (from $10^{-7}$ to $10^{-13}$; most opaque corresponds to $p_x = 10^{-13}$). The dotted diagonal lines indicate the LER$_X$ floor.
At moderate target LER the two schedules have comparable overhead; the simultaneous schedule becomes advantageous only in the regime of very low target LER where its higher $\Lambda$ compensates the extra qubits.

\section{Compiling LDPC parity checks with QND multi-Z measurements}
\label{sec:ldpc_parity}

In the context of high noise-bias, it is advantageous to implement a more efficient phase-flip code than the repetition code \cite{Ruiz-LDPC-2025}.
In order to do this, one needs to perform four-body \(M_{X^{\otimes 4}}\) stabilizer measurements. Note that these same weight-4 stabilizer measurements also enable a significant hardware cost reduction for magic state preparation with biased-noise qubits~\cite{Ruiz-npj-2026}. This measurement can clearly be compiled using CNOT gates implemented via QND \(M_{Z^{\otimes 3}}\) measurements.
It is also direct to generalize the circuit in Fig~\ref{fig:CX}(b) to perform a \(M_{X^{\otimes k}}\) measurement with a single QND \(M_{Z^{\otimes 2k}}\) followed by \(k\) single-qubit \(M_{X}\).
Although it seems unlikely that for \(k>2\) this could be implemented efficiently in hardware.
We show here that we can use only QND \(M_{Z^{\otimes 4}}\) measurements with the same number of auxilliary qubits.
In Figure~\ref{fig:MX4}, we present this hardware-efficient version for a \(M_{X^{\otimes 4}}\) compiled with QND \(M_{Z^{\otimes 4}}\) measurements.
One needs 4 auxiliary qubits prepared in \(\ket{+}\) and to perfom 3 \(M_{Z^{\otimes 4}}\) overlapping on pairs of data and auxiliary qubits.
Then single qubit \(M_{X}\) on the initial data qubits teleporting them to the auxillary ones together with perfoming the desired \(M_{X^{\otimes 4}}\) measurement whose outcome is the product of the individual \(M_X\) outcomes.
The \(Z\) correction to correctly teleport the data qubits are represented in the circuit of Figure~\ref{fig:MX4}.
The \(X\) correction is found by matching the non-trivial outcomes of the QND \(M_{Z^{\otimes 4}}\) measurements.
Note that one can perform more QND \(M_{Z^{\otimes 4}}\) measurements which are redundant and would enable detection or even correction of a measurement error or incoming \(X\) errors.
For instance the grey \(M_{Z^{\otimes 4}}\) in Figure~\ref{fig:MX4} enables detection of one \(X\) error.

\end{document}